# *GenomeFingerprinter* and universal genome fingerprint analysis for systematic comparative genomics

**Yuncan Ai\*, Hannan Ai, Fanmei Meng, Lei Zhao**

State Key Laboratory for Biocontrol, School of Life Sciences, Sun Yat-sen University, Guangzhou 510275, P. R. China

Lssayc@mail.sysu.edu.cn

\* Corresponding Author:

    Yuncan Ai

    Ph.D., Professor, Principal Investigator

    Lssayc@mail.sysu.edu.cn

    State Key Laboratory for Biocontrol

    School of Life Sciences

    Sun Yat-sen University

    Guangzhou 510275

    P. R. China

**Abstract**

**Background:**

How to compare whole genome sequences at large scale has not been achieved via conventional methods based on pair-wisely base-to-base comparison; nevertheless, no attention was paid to handle in-one-sitting a number of genomes crossing genetic category (chromosome, plasmid, and phage) with farther divergences (much less or no homologous) over large size ranges (from Kbp to Mbp). It should be a priority to persue comparative genomics at large scale based on geometrical analysis of sequence in the post-genomic era. However, even how to simply visualize a DNA sequence has been challenging for decades; little progress has been made to date.

**Results:**

We created a new method, *GenomeFingerprinter*, to unambiguously produce three-dimensional coordinates from a sequence, followed by one three-dimensional plot and six two-dimensional trajectory projections to illustrate whole genome fingerprints. We further developed a set of concepts and tools and thereby established a new method, universal genome fingerprint analysis. We demonstrated their applications through case studies on over a hundred of genome sequences. Particularly, we defined the total genetic component configuration (TGCC) (i.e., chromosome, plasmid, and phage) for describing a strain as a system, and the universal genome fingerprint map (UGFM) of TGCC for differentiating a strain as a universal system, as well as the systematic comparative genomics (SCG) for comparing in-one-sitting a number of genomes crossing genetic category in diverse strains. By using UGFM (I), UGFM-TGCC (II), and UGFM-TGCC-SCG (III), we compared a number of genome sequences with farther divergences (chromosome, plasmid, and phage; bacterium, archaeal bacterium, and virus) over large size ranges (6Kbp~5Mbp), giving new insights into critical problematic issues in microbial genomics in the post-genomic era.

**Conclusion:**

This paper provided a new method for rapidly computing, geometrically visualizing, and intuitively comparing genome sequences at fingerprint level, and hence established a new method of universal genome fingerprint analysis for systematic comparative genomics.

**Introduction**

By using conventional methods based on pair-wisely base-to-base comparison, how to compare whole genome sequences at large scale has not been achieved; nevertheless, no attention was paid to handle in-one-sitting a number of genomes crossing different genetic category (chromosome, plasmid, and phage) with farther divergences (less or no homologous among genetic components) over large size ranges (from Kbp to Mbp per sequence). It should be a priority to persue comparative genomics at large scale based on geometrical analysis of sequence in the post-genomic era. However, little progress has been made to date; even how to simply visualize a DNA sequence has been challenging for decades [1].

Pioneering works in computer reading and geometrical visualizing of DNA sequence had been tried in one-dimension [2, 3], two-dimensions (Z-curve) [4], and three-dimensions (H-curve) [5, 6]. However, those were valid only for 'static' modeling and visualizing. The 'dynamic' modeling and visualizing in a virtual reality environment had been studied [7, 8]. A comprehensive example was AND-viewer, which provided a three-dimensional way to dynamically sense the big picture of a large DNA sequence in a virtual reality environment by using sensor, instead of mouse or keyboard. This pioneering effort had made fantastic progress for human to mimic 3D visions to intuitively sense genome sequences [7, 8], but still there was no possibility of using the datasets created for visions to further explore real biological contexts.

The post-genomic era promoted demanding of data mining and robust reasoning with huge amount of genome sequences [1]. Comparative genomics was essential to retrieve and mine genome sequences, there were numerous conventional methods, which were divided into two types: algebraic approach [9, 10, 11, 12] and geometrical approach [13, 14].

Algebraic approach means that calculating dissimilarity, or similarity or identity are based on pair-wisely base-to-base comparison; the outputs of calculation are only used for visualization through graphic techniques, rather than for robust data mining and reasoning analysis. Of course, tools for genomic data visualization are still essential for end-users to explore, interpret, and manipulate data [1]. The most common tools were BLAST [9] and CLUSTALW [10], which were only for pair-wisely comparisons among a certain number of short fragments at gene level. Recently, a BLAST-based visualization tool, BRIG, was constructed for genome-wide comparison to create images of multiple circular genomes among a number of closely related bacteria strains [11]. The output image showed BLAST-similarity between a central reference sequence and other sequences in question as a set of concentric rings, where BLAST-matches were colored on a sliding scale indicating a defined percentage of BLAST-identity. It had great advantage over other common tools, like ACT [12], in terms of the numbers of genomes being handled simultaneously and the ways of comparing and presenting of images in-one-sitting. These features made it a versatile approach for visualizing a range of genomic data, but it still was only for visions. Mauve [14, 15] was widely used for comparing and visualizing a number of genomes of close relatives in linear forms, which combined algebraic calculation and graphic display. However, even with close relatives only, the number of genomes being calculated and displayed dramatically depended on computation constraints causing too much CPU time (at least $O(n^2)$ in time complexity) or memory overflow, which limited to a fewer genomes of close relatives to be compared at once time.

Geometrical approach means that both calculation and visualization are dynamic for geometrical analysis with the input and output re-useable. One promising example was Zplotter (in Z-curve method) calculating three-dimensional coordinates from a linear genome sequence. Those coordinates were used to create a three-dimensional geometrical vision in a rough manner (as open Z-curve) for a given DNA sequence [16]. Hundreds of such visions for microbial genomes were collected as a database [17]. The Z-curve method (based on Zplotter) was not only used for simple visualization [16, 17], but also for geometrical analysis to further explore real contexts of biology [18, 19, 20, 21]. For example, two replication *ori* points in archaeal genomes were predicted by Z-curve method [22, 23] and confirmed later by wet experiments in other labs [24, 25], showing its promising. Z-curve method was widely used by researchers around the world and had promoted the progress in understanding of genomics, starting a new frontier in geometrical analysis of genome sequences. However, Zplotter algorithm had an inevitable limitation in mispresentation of a genome sequence but with different cutting-points (explanations in the main text), which would not be suitable for creating unique genome fingerprints.

In this paper, we present a new method, called *GenomeFingerprinter*, to unambiguously produce three-dimensional coordinates from sequence, followed by one three-dimensional plot and six two-dimensional trajectory projections to illustrate whole genome fingerprints. We further develop a set of concepts and tools and thereby establish a new method called universal genome fingerprint analysis. We demonstrate their applications through case studies on over a hundred of genome sequences, giving new insights into critical problematic issues in microbial comparative genomics. We anticipate that our methods could be widely applicable to systematic comparative genomics in the post-genomic era [1].

**Results**

**Mathematical model and three-dimensional coordinates**

For geometrical visualization of a given genome sequence, the key step is how to get its three dimensional coordinates ($x_n$, $y_n$, $z_n$). To do this, the Z-curve [16] defined a coordinate ($x_n$, $y_n$, $z_n$) for each base in a linear genome sequence (n=1, 2, ..., N; N is the sequence length) by the equation (**0**). It defined a unique Z-curve from a given linear sequence, and *vice versa*. Note, it was designed for a linear genome sequence and $A_n$, $T_n$, $G_n$, $C_n$ were the sum of numbers for each of four base-type (A, T, G, C), respectively, counting from the first base to other bases before and including the $n^{th}$ base in a linear sequence (n=1, 2, ..., N). The calculations could be performed by using Zplotter program [16]. The main problem here was the ambiguousness of the "first base" due to cutting-point errors in deposited genome sequences (see explanations later).

$$\begin{cases} x_n = (A_n + G_n) - (C_n + T_n) \\ y_n = (A_n + C_n) - (G_n + T_n) \\ z_n = (A_n + T_n) - (C_n + G_n) \end{cases} \quad (\mathbf{0})$$

Here, we take the same definition as equation (**0**), but with different contents of $A_n$, $T_n$, $G_n$, $C_n$. To do that, we thus propose a new mathematic model, called *GenomeFingerprinter*, for geometrical visualization of a circular genome sequence. A circular sequence contains 40-bps (5'-3'): ACACTGACGCACACTGACGCACACTGACGCACACTGACGC (Figure 1) as an artificial example will be used to illustrate the conceptual principals of our method.

Firstly, we randomly select a base ($n^{th}$) as the first target base (TB). For the given TB ($n^{th}$), we define its relative distances (RD) (**1**) to the other moving $m^{th}$ base (as focusing base, FB) (m=1, 2, ..., N).

$$RD_n^m = \begin{cases} 1, & (m = n+1) \\ 2, & (m = n+2) \\ ... & ... \\ N-1, & (m = n+n-1) \\ N, & (m = n+n) \end{cases} \quad (1)$$

Here, the RD concept is critical. Once we have selected the given TB (suppose at position 1, base **A**) and the other moving FB (suppose at position 20, base **C**), the RD value is 19 (Figure 1). Thus, a collection of RD values (m=1, 2, …, N) will be generated for the given TB. Particularly, the RD formula (**1**) can virtually treat an arbitrary linear sequence as a circular one. For example, when the moving $m^{th}$ FB is located at the position of n+n, the RD is N, which means the RD value now is N, not zero, when the moving $m^{th}$ FB going over one circle (i.e., starting from the position at the $n^{th}$ base and finishing at the same position at the $n^{th}$ base).

Secondly, we define the weighted relative distance (WRD) (**2**) (N is the sequence length). The example above will have value 19/40. This is simply to reduce memory burden and thus release computation constraints for larger sequences.

$$WRD_n^m = \frac{RD_n^m}{N} \quad (2)$$

Thirdly, for the same chosen TB ($n^{th}$), we define the sum of the weighted relative distances (SWRD) (**3**) from the above collection of WRD (m=1, 2, …, N) for each of the four base-type (A, G, T, C), respectively.

$$\begin{aligned} SWRD_n^A &= \sum_n^A [WRD_n^m] \\ SWRD_n^G &= \sum_n^G [WRD_n^m] \\ SWRD_n^T &= \sum_n^T [WRD_n^m] \\ SWRD_n^C &= \sum_n^C [WRD_n^m] \end{aligned} \quad (3)$$

Fourthly, we define the coordinate ($x_n$, $y_n$, $z_n$) (**4**) for the chosen TB ($n^{th}$). Note, here we count the sum of the weighted relative distances (SWRD) (unlike Z-curve method counting the sum of numbers) for each of four base-type (A, T, G, C), respectively. So far, that is only one cycle done for only one chosen TB ($n^{th}$).

$$\begin{cases} x_n = (SWRD_n^A + SWRD_n^G) - (SWRD_n^C + SWRD_n^T) \\ y_n = (SWRD_n^A + SWRD_n^C) - (SWRD_n^G + SWRD_n^T) \\ z_n = (SWRD_n^A + SWRD_n^T) - (SWRD_n^C + SWRD_n^G) \end{cases} \quad (4)$$

Finally, we will repeat the above cycle, i.e., selecting the next TB (e.g., n=2 here) and repeating the process. We will have total N cycles (n=1, 2, …, N); each cycle has only one chosen TB and creates only one coordinate ($x_n$, $y_n$, $z_n$) for that chosen TB. All N bases will have their coordinates ($x_n$, $y_n$, $z_n$) after having finished all of these N cycles. We have developed in-house script, GenomeFingerprinter.exe, to do all.

As an example, the artificial 40-bps genome sequence (Figure 1) had its coordinates ($x_n$, $y_n$, $z_n$) (Table 1), which were calculated by using our program GenomeFingerprinter.exe, giving each base with a coordinate ($x_n$, $y_n$, $z_n$) as a point in the three-dimensional space, in total 40 points for the whole sequence.

**Three-dimensional plot (3D-P) and primary genome fingerprint map (P-GFM)**

The three-dimensional coordinates ($x_n$, $y_n$, $z_n$) can be plotted out as a three-dimensional plot (3D-P) to give a geometrical vision. The artificial 40-bps sequence had only 40 points (Table 1) hence giving a naive vision. As real examples, we showed visions for fragmental sequences ranging from tens to hundreds of kilobases (Table 2) of *Escherichia coli* strains (Figure 2). Clearly, each vision had its unique genome fingerprint (GF) both globally and locally. We defined such a GF vision as genome fingerprint map (GFM). The GFM was an intuitive identity for an individual genome sequence, and vice versa. Therefore, from now on, we can directly operate and compare GFM for studying sequence. That is, we compare genome sequences through genome fingerprints (via geometrical analysis) instead of sequence base-pairs (via algebraic analysis). For convenience, we further defined the three-dimensional plot vision as the primary genome fingerprint map (P-GFM).

**Two-dimensional trajectory projections (2D-TPs) and secondary genome fingerprint maps (S-GFMs)**

To demonstrate sophisticated genome fingerprints, we created six two-dimensional trajectory projections (2D-TPs) for a given P-GFM by combining different components from its coordinates, including $x_n \sim n$, $y_n \sim n$, $z_n \sim n$, $x_n \sim y_n$, $x_n \sim z_n$, and $y_n \sim z_n$. For convenience, we defined these six 2D-TPs as the secondary genome fingerprint maps (S-GFMs). For example, six S-GFMs of *Escherichia coli* K-12/W3110 genome sequence clearly showed subtle variations both globally and locally (Figure 3). Note that S-GFMs of $x_n \sim z_n$, $y_n \sim z_n$, $x_n \sim y_n$ beared much more sensitive information compared to those of $x_n \sim n$, $y_n \sim n$, and $z_n \sim n$, respectively. Generally, S-GFMs can amplify subtle variations that are insensitive or invisible in P-GFMs. Particularly, S-GFMs of $x_n \sim y_n$, $x_n \sim z_n$ and $y_n \sim z_n$ are much more sensitive in differentiating local subtle variations, intuitively identifying unique genome features; whereas S-GFMs of $x_n \sim n$, $y_n \sim n$ and $z_n \sim n$ are relatively less informative but useful when focusing on global patterns (Figure 3).

**Universal genome fingerprint map (UGFM)**

P-GFM and S-GFMs can be either separately or sequentially used. For convenience, we defined the universal genome fingerprint map (UGFM) to unify both of them. By UGFM, we could compare in-one-sitting a number of sequences and display their GFMs at once time, on which each GFM could be classified into different groups solely based on its location (Figure 4).

For example, six archaeal genomes and twelve fragmental sequences from *E.coli* strains (Table 2) had complex P-GFMs (Figure 4). Within the species *Sulfolobus* *islandicus*, strains M.14.25 and M.16.4 shared global similarity in P-GFM (Figure 4, A), indicating subtle variations at strain level. However, with farther divergences, strain *S. islandicus* Y.N.15.51 globally differed from *Methanococcus voltae* A3 but locally shared similar regions in P-GFM (Figure 4, B); whereas *S. islandicus* Y.G.57.14 completely differed from *Methanosphaera stadtmanae* 3091 (Figure 4, C), confirming their farther lineages.

On the other hand, within the species *Sulfolobus islandicus*, two strains M.14.25 and M.16.4 had only subtle variations (Figure 4, A), how could they be precisely differentiated by P-GFM? We defined geometrical center ($\bar{x}, \bar{y}, \bar{z}$) as a distinctive indicator for a single P-GFM to compare individual P-GFMs. For example, two strains M.14.25 and M.16.4 had different geometrical center values (644.00, -2081.00, 388729.14) and (476.50, -1916.50, 387938.64), respectively, and hence were clearly distinguishable.

Furthermore, those twelve fragmental sequences from *E.coli* strains (Table 2, Figure 4, D) were further enlarged and displayed as a UGFM besides their own individual P-GFMs (Figure 5). Clearly, there were six groups on UGFM (Figure 5, A, B, C, D, E, F) solely based on the locations of different P-GFMs. Particularly, different fragmental sequences either from the same strain (e.g., 91.1.1, 91.1.61, 91.6.59) or from different strains (e.g., 913.5.57, 4431.1.70, 7946.4.7, 10473.1.74, 10498.4.86, 12947.1.50, 13941.2.60) (Table 2) could be revealed as complicated P-GFM patterns. Some were similar (91.1.61, 913.1.77 and 10473.1.74 (Figure 5, A); 91.6.59, 913.5.57 and 13941.2.60 (Figure 5, B) ) but most were different (Figure 5, C, D, E, F) no matter what the lineage was, strongly demonstrating the facts that there were modular domains in these genomes and such mosaic structures probably remained their tracking of evolutionary history. Most interestingly, a given P-GFM had quite different views between its own P-GFM and that on UGFM simply because of the scale-down and view-angle rotation effect in UGFM (Figure 5). This feature would ensure UGFM as a powerful tool for large-scale global comparison in-one-sitting among a large number of whole genome sequences, theoretically, as many sequences as possible as long as the computer memory could allow.

**Universal genome fingerprint analysis (UGFA)**

Now that we had such a powerful tool, UGFM, based on unambiguous genome fingerprints, to compare a large number of whole genome sequences in-one-sitting (Figure 5), we further established a new method called universal genome fingerprint analysis (UGFA) (Figure 6). We anticipated that UGFA would be effective for systematic comparative genomics at large scale by expanding the scope of genetic category in question. Briefly, the UGFA method consisted of three subcategories (Figure 6): UGFM (I) (Figure 7, 8, 9), UGFM-TGCC (II) (Figure 10), and UGFM-TGCC-SCG (III) (Figure 11) corresponding to three objects: a genome, a strain, and a set of strains, respectively. For each subcategory, demonstration with examples of case studies was described in details below.

*UGFM (I): Universal genome fingerprint map (UGFM)*

Firstly, UGFM (I) was the foundation and the first major component of our UGFA method. It was proved powerful in global comparison at large scale for prokaryote bacteria genomes (Figure 5). More examples were from a number of genomes of archaeal bacterium (Table 2, Figure 7), phage (Table 3, Figure 8), and virus (Table 3, Figure 9). Five archaeal bacteria strains (*Halomonas elongate* DSM 2581, *Halorhodospira halophilia* SL1, *Halorhabdus utahensis* DSM 12940, *Halothermothrix orenii* H 168 and *Halothiobacillus neapolitanus* c2) representing five genera of halophilic Archaea were displayed as a UGFM (Figure 7) showing larger scale-down and view-angle rotation effect. Clearly, each strain had only one chromosome with size ranging of 2.6 ~4.1 Mbp; and these five archaeal chromosomes had no close relationships at all (Figure 7) confirming their farther diverse lineages at genus level (Table 2,). However, forty seven phages of family *Microviridae* (Table 3) that were grouped into two major clusters (Figure 8) and twenty four coronavirus strains (Table 3) that were classified into seven clusters (Figure 9) perfectly matched to their biological identities among close relatives. Put together, these fingdings from total eight three genomes (i.e., twelve bacteria, five archaeal bacteria, forty seven phages, and twenty four viruses) as good examples demonstrated that UGFM (I) could apply to any genetic category (bacterium, archaeal bacterium, phage, and virus) no matter how farther (Figure 7) or closer (Figure 5, 8, 9) divergences of genetic components in comparison.

### *UGFM-TGCC (II): Universal genome fingerprint map (UGFM) of total genetic component configuration (TGCC)*

Secondly, how to compare a number of genome sequences crossing different genetic category (e.g., chromosome, plasmid, and phage) in a strain? Accordingly, we defined the total genetic component configuration (TGCC) as a set of genomes crossing all genetic category (chromosome, plasmid, and phage, if applicable) in a strain for describing a strain as a system. We further defined the universal genome fingerprint map (UGFM) of total genetic component configuration (TGCC) (UGFM-TGCC) for differentiating a strain in view of a universal system. Therefore, we could use UGFM-TGCC (II) to compare in-one-sitting among all genetic components in a strain, regardless the format of category (chromosome, plasmid, and phage).

For example, four strains crossing four genera in haophilic Archaea (Table 2) including *Halogeometricum boringquense* DSM 11551 (one chromosome and five plasmids), *Haloterrigena turkmenica* DSM 5511 (one chromosome and six plasmids), *Natrialba magadii* ATCC 43099 (one chromosome and three plasmids), *Natrinema pellirubrum* DSM 15624 (one chromosome and two plasmids) were demonstrated by UGFM-TGCC (II) (Figure 10) clearly indicating their farther lineages at genus level. Note that the scale-down and view-angle rotation effect revealed the farther divergences between one chromosome and multiple plasmids in a certain strain, suggesting that it would be challenging for conventional methods to compare them due to much less or no homologous. Specifically, in the same figure (Figure 10, H), the tiny green spot (plasmid NC_008213) and the giant red vision (chromosome NC_008212) with farther divergences would not be easily compared by any other conventional methods.

***UGFM-TGCC-SCG (III): UGFM-TGCC-based systematic comparative genomics (SCG)***

Thirdly, how to compare a number of genome sequences both crossing genetic category in a strain (chromosome, plasmid, and phage) and crossing a number of strains (a cluster of strains) as a system (i.e., in-one-sitting)? To compare a number of such diverse genomes in-one-sitting, we defined a concept of UGFM-TGCC-SCG (III), UGFM-TGCC-based systematic comparative genomics (SCG). Note, here we called it as "systematic comparative genomics (SCG)" simply because all genomes crossing different genetic category (chromosome, plasmid, and phage) among diverse strains should be much less or even no homologous at all, which would be incredibly challenging to any known conventional methods that principally based on similarity analysis of homologous. In other words, to our knowledge to date, no one conventional method could handle such farther diverse genetic components in-one-sitting; to which even no attention was paid before. In fact, all of the documented researches on comparative genomics to date were automatically based on the assumption that there was so-called a reference genome sequence for very close relatives in question; otherwise, they would not bother to do comparison. But, in our case, we exactly focused on the opposites that had much less or even no homologous and compared those diverse genetic components crossing farther divergences regardless the format of genetic category and regardless the extent of lineage divergence. Therefore, we called our objects in comparison as the "systematic comparative genomics" in order to distinguish from other traditional routes. This was one of the core concepts and aims in the present study.

Indeed, our UGFM-TGCC-SCG (III) subcategory method was powerful to handle those extraordinary situations. For example, total nineteen genomes including six chromosomes and thirteen plasmids with larger size ranges (6Kbp ~ 4Mbp) could be, separately, mapped and analyzed by using UGFM-TGCC-SCG (III) (Figure 11). These nineteen genomes were from four strains crossing four genera of halophilic Archaea (Table 2) and analyzed in-one-sitting as two sets of comparison (Figure 11): *Halorubrum lacusprofundii* ATCC 49239 (two chromosomes and one plasmid) *vs.* *Haloarcula marismortui* ATCC 43049 (two chromosomes and seven plasmids); *Haloferax vocanii* DS2 (one chromosome and four plasmids) *vs.* *Halomicrobium mukohataei* DSM 12286 (one chromosome and one plasmid). Obviously, they were certainly demonstrated as diverse lineages solely based on genome fingerprints. Most importantly, note that tiny spots (e.g., corresponding to 6Kbp) and giant ones (e.g., corresponding to 4Mbp) were harmoniously existed in the same figures (Figure 10, H, Figure 11, C), either closely or farther away. Such amazing landscapes could be only revealed by our unique methods under the notions of so-called "total genetic component configuration" and "systematic comparative genomics", particularly, as UGFM-TGCC and UGFM-TGCC-SCG in these cases. These should be more than enough as representatives to prove UGFM-TGCC-SCG (III) effective and powerful.

**Case studies: Applications of universal genome fingerprint analysis (UGFA)**

*Objectives*

As for more specific examples, we chose two archaeal halobacteria strains, *Halobacterium* sp. NRC-1 and *Halobacterium salinarum* R1 (Table 2) [26, 27], to persue systematic comparative genomics by using our UGFA method. It was not only because they had incredible microbiological features such as genome-wide evolution events [28, 29] and multiple replication *ori* points (unlike the common prokaryotes with only one replication *ori* point) that could be easily tracking [23], but also because two genomes were independently sequenced by two labs [28, 29] and had led to interesting arguments about critical problematic issues in microbial genomics and taxonomy, such as whether they were the same species or strain [29] or their genome sequences were correctly assembled particularly considering of mega-plasmids or minichromosomes [30, 31], and what might be the mechanism for evolutions [29, 31, 32, 33], even what should be considered for refining a "species" in taxonomy [34]. We expected that the universal genome fingerprint analysis (UGFA) could provide new insights into these critical problematic issues that would be crucial and invaluable for modern microbiology in the post-genomic era.

*Genome-wide evolution events*

From the sophisticated genome fingerprints in S-GFMs (Figure 12), two deposited chromosomes NRC-1 (NC_002607) and R1 (NC_010364) were very similar but not identical having subtle differences at strain level (Figure 12, A, B, E, F), supporting the claim that two strains were virtually from the same ancestor but had undergone evolutions [31, 32]. The subtle variations (Figure 12, C, D, G) also indicated the genome-wide evolution events (shown by arrow-markers), causing longer of NRC-1 chromosome. It was coincided with the documented facts that the IS-element-rich regions [27] shuttled between chromosomes and mega-plasmids [30, 31], but the core genes conserved [33].

*Two replication ori points*

Again, from S-GFMs, two replication *ori* points, *ori*C1 and *ori*C2, (Figure 12, E, F, H) were identified and the replication domains in two genomes were demonstrated identical, and those evolution events were not located in such replication regions. These evidences also supported the claim that two strains virtually came from the same ancestor strain [31, 32]. Most interestingly, two replication *ori* points in strain NRC-1 were reported as the first representatives of archaeal bacteria, changing the traditional definition of only one *ori* point in prokaryotes. In fact, one of two *ori* points was predicted by theoretical Z-curve analysis [22, 23] and confirmed by biological experiments later [24, 25]. Thus the reproducibility in identification of such two replication *ori* points in two sequences has proved that our method is as effective and sensitive as Z-curve analysis [22, 23].

*UGFM-TGCC-SCG for differentiating strains*

Two strains NRC-1 and R1 were completely different in terms of the numbers of plasmids and total base-pairs (Table 4). How to concisely describe their differences in visualization remained challenging. For example, they had eight genetic components including two chromosomes and six plasmids (Table 4), which made it ambiguous to only compare any part of them as traditional comparative genomics did. We thus should compare all of eight genomes in order to differentiate two strains unambiguously. By using UGFM-TGCC-SCG (Figure 13), we compared eight genome sequences with farther divergences crossing different genetic category (i.e., chromosome and plasmid) over large size ranges (40Kb ~ 3Mb) (Table 4) that would be challenging for conventional methods. In short, the UGFM-TGCC-SCG vision clearly confirmed that two strains were completely different and eight components had farther divergences (Figure 13, B). Particularly, two chromosomes were almost the same (Figure 13, B) but six plasmids had larger divergences (Figure 13, A), strongly indicating that plasmids had no close lineages with chromosomes in two strains. In addition, two mega-plasmids in strain NRC-1 had no close lineages with four plasmids in strain R1, suggesting there was no possibility to misassemble them due to less homologous. Most interestingly, even within the same strain, chromosome and plasmid showed distinctive lineage divergences. In other words, there was no correlation between chromosome and plasmid within a certain strain (i.e., no binding to a certain strain), indicating possible independent evolution among chromosomes and plasmids.

### *Double-check between UGFM-TGCC-SCG and Mauve*

To double-check the lineages revealed by UGFM-TGCC-SCG (Figure 13), we used progressiveMauve mode [14] analysis to make pair-wisely multiple genome alignment among eight components (Figure 14, A) showing overall bare homologous although it took much longer time; whereas Mauve mode [15] analysis failed in such a comparison because it stopped alignment due to no essential homologous, as we predicted beforehand. Mauve mode [15] analysis yet worked well, separately, with subsets of six plasmids (Figure 14, B) and two chromosomes (Figure 14, C), respectively, and confirmed those partial relationships among six plasmids and between two chromosomes. We thus concluded that the lineage relationships revealed by UGFM-TGCC-SCG could be partially confirmed by Mauve method and confirmed that two strains were completely different and remained as sister-strains within one species. In other words, in this case, progressiveMauve mode [14] could barely compatible to UGFM-TGCC-SCG (III) whereas Mauve mode [15] did not, but it could be used to deal with subsets, separately.

**Discussion**

As mentioned before, how to compare whole genome sequences at large scale has not been achieved by using conventional methods [11, 14] that based on pair-wisely base-to-base sequence similarity analysis; even no attention was paid to handle in-one-sitting a number of genomes crossing different genetic category with farther divergences (e.g., less or no homologous among crossing genetic components: chromosome, plasmid, and phage; bacterium, archaeal bacterium, and virus) over large size ranges (e.g., from Kbp to Mbp per genome sequence). We believe that how to persue comparative genomics at large scale based on geometrical analysis of sequence, rather than pair-wisely base-to-base comparison, will be a priority in the post-genomic era. However, little progress has been made to date, even how to visualize a DNA sequence has been challenging for decades [1]. To our knowledge to date, no method for creating "unambiguous genome fingerprint (GF)" was documented; no concept of "universal genome fingerprint analysis (UGFA)", or "total genetic component configuration (TGCC)", or "systematic comparative genomics (SCG)" was proposed. Particularly, note that all sequences of components both crossing different genetic category (e.g., chromosome, plasmid, and phage; bacterium, archaeal bacterium, and virus) and crossing a number of diverse strains in-one-sitting should be much less or no homologous at all, which would be incredibly challenging to any known conventional methods that principally based on pair-wisely base-to-base homologous analysis. No conventional method could handle in-one-sitting such farther diverse genetic components. Therefore, it would be impossible to compare our methods, *GenomeFingerprinter* and universal genome fingerprint analysis (UGFA), as a whole system with other documented methods in terms of advantages and disadvantages. However, in the present study, we tried best to compare partial features with two programs partly related to ours.

### *GenomeFingerprinter vs.* Zplotter

#### *Validity*

Zplotter (in Z-curve method as a geometrical-type approach) was mainly used to create coordinates for subsequent use by Z-curve analysis, but not used for what we proposed as creating the "genome fingerprint (GF)" and the "universal genome fingerprint analysis (UGFA)" in the present study. Although Zplotter's coordinates were used to produce hundreds of graphs (as Z-curves) of microbial genomes documented as a database [17], there were no stable features in terms of so-called fingerprints. In fact, for example, we re-plotted visions for *Halobacterium* sp. NRC-1 genome sequence [NC_002607] by using Zplotter's coordinates of either $z_n$' (Figure 15, B) or $z_n$ (Figure 15, C) to present as an open rough Z-curve. Note that those visions themselves created by using $z_n$' and $z_n$, respectively, were quite different from each other due to wavelet transform in the algorithm of Zplotter [16]. In contrast, our method presented a unique circular vision with accurate and delicate fingerprint for the same genome (Figure 15, A). Also note that using $z_n$' (Figure 15, B) showed a similar frame of vision to ours except that it was in an open rough Z-curve with lesser features whereas using $z_n$ (Figure 15, C) gave a complete different vision from ours. We thus recommend that our *GenomeFingerprinter* method could be an alternate of Zplotter to provide more accurate and delicate coordinates for Z-curve analysis, but should be aware of choosing whether $z_n$ from our method or $z_n$' from Zplotter, referring to the specific questions for various researches.

*Reliability*

Furthermore, we had found a major problem when using Zplotter to handle circular genome sequences with cutting-point errors. In fact, Zplotter was designed for a linear sequence [16] because its algorithm depends on counting the absolute numbers of bases starting from the "first" base in a given linear sequence. In fact, when a deposited sequence as a linear form (i.e., no matter what the original form should be as either linear or circular), even the same circular sequence with cutting-point errors changing its real "first" base could be quite different for the input to Zplotter so that the output visions were differently presented (Figure 15, B, C). In contrast, our method was initially created for a circular sequence (Figure 1), but it could apply to a linear form since linear one would be a specific form of circular one and particularly because our method measured the relative distance in a circular form (as discussed with the formula (**1**) before, Figure 1), rather than the absolute numbers of bases counting from the "first" base in a linear sequence. For example, the same circular sequence *Halobacterium* sp. NRC-1 (NC_002607) with different cutting-point (e.g., NC_002607_RC re-cut at 700 kbps) were incorrectly presented as different visions by using Zplotter's coordinates (Figure 15, B, C), whereas the exact same vision was shown by our method (Figure 15, A). Thus our method was valid for both circular and linear forms and no matter where the cutting-point was.

*Adaptability*

Finally, we would like to address the fundamental scientific principals for why dealing with circular genomes should be critical for microbes. That was overlooked in literatures before.

Theoretically, the circular form [32] would be much more stable than its linear form in living cells. In nature, most microbial genomes are in circular double strands, which protect them from natural degradation because of relatively simple structure. Also, the circular genomes and their linear forms are usually changing into each other only when they are living at certain functioning stages, such as rolling-model replication and plasmid-mediated conjunction. Most importantly, circular and linear forms are functioning both genetically and physiologically in a coordinated way for a given genome in a given microbe. In other words, their forms are changeable into each other only when responding to real living conditions [32, 33]. Anyway, we could catch up the circular form status in life cycles.

Technically, the techniques and people in different groups were not yet unified to guarantee all deposited genome sequences in correct forms. In fact, most sequences deposited in public databases so far were neither in their natural orders of starting from the real "first" base, nor in the direction from 5' to 3'. We thus had to tackle with such cutting-point errors, as illustrated by examples (Figure 15). Fortunately, as mentioned before, the RD formula (**1**) in our method could virtually treat an arbitrary linear sequence as a circular one (Figure 1), avoiding the impact of any possible cutting-point errors existed in public deposited sequences.

Informatively, the closed (or in circular form) fingerprint beared much more information, concerning with genome-wide comparative genomics at fingerprint level (Figure 12, 13). Most importantly, our method was initially designed for circular forms (Figure 1), but finally was proved not ambiguous for linear forms when dealing with cutting-point errors (Figure 15). In other words, our method could precisely calculate the three-dimensional coordinates for a given circular or linear sequence with or without correct cutting-point, could accordingly present a unique genome fingerprint giving a certain geometrical center ($\overline{x}$, $\overline{y}$, $\overline{z}$) (Figure 15), and could consequently guarantee the subsequent unambiguous trajectory projections. In short, our method guaranteed the validity of universal genome fingerprint analysis.

To sum up, we conclude that *GenomeFingerprinter* has advantages over Zplotter in creating unambiguous coordinates and therefore can be an alternate component of Z-curve method, which can be widely applicable to all aspects established by Z-curve method to date [18, 19, 20, 21, 22, 23] and beyond.

### *GenomeFingerprinter vs.* Mauve

#### *Efficiency*

Mauve (as an typical algebraic-type program), combined computing and plotting in-one-sitting, is commonly used for pair-wisely comparison and vision [14]. However, it had difficulty with a number of larger genome sequences due to its inner constraints, either too slow or memory overflow (MO). In contrast, our method could rapidly calculate and visualize, separately, tens of large genomes. For example, Mauve had at least $O(n^2)$ whereas our method had $O(n)$ in time complexity (Table 5). By using our method, only if plotting all larger graphics in-one-sitting would cause memory overflow. Examples were five bacterial chromosome genomes (Figure 7), forty seven phage genomes (Figure 8), and twenty four virus genomes (Figure 9), respectively, that could be easily plotted out in-one-sitting. Particularly, our method calculated and visualized, separately, and thus not only ensured the higher performance efficiency for large set of genomes (Table 5) but also offered both inputs and outputs re-usable for the subsequent processes of universal genome fingerprint analysis and beyond (e.g., for Z-curve analysis consequently).

On the other hand, Mauve had two modes: progressiveMauve mode [14] and Mauve mode [15]. As discussed before, only progressiveMauve mode [14] could partially deal with what we so-called systematic comparative genomics (Figure 14, A) showing overall bare homologous although it took much longer time; whereas Mauve mode [15] failed in the comparison because it stopped alignment due to no essential homologous. Mauve mode [15] analysis yet worked well, separately, with subsets of six plasmids (Figure 14, B) and two chromosomes (Figure 14, C), respectively, and confirmed those partial relationships among six plasmids and between two chromosomes.

*Prediction*

Mauve [14, 15] can visualize what it is, but can not predict what it should be without a reference sequence or specific pre-knowledge. In contrast, our method provides geometrical analysis of genome fingerprints with six trajectory projections, which intuitively predict unique features such as genome-wide evolution events and replication *ori* points (Figure 12), either based on a reference sequence or derived from common knowledge.

*Compatibility*

From universal genome fingerprint analysis (UGFA), the subtle variations (Figure 12, C, D, G) could predict genome-wide evolution events at small scale in chromosomes, but no direct evidence yet could be drawn. Thus we used Mauve to pair-wisely compare two genomes and confirmed genome-wide evolution events (Figure 12, C), demonstrating that our method can rapidly predict evolution events and Mauve can precisely test and confirm such predictions by showing out the predicted specific regions (Figure 12, C). The same was true for UGFM-TGCC-SCG by our method (Figure 13) and the pair-wisely comparison by progressiveMauve (Figure 14, A). Thus, we recommend that our method and Mauve method are compatible and partners, taking both advantages of our method for rapid and intuitive prediction in general and Mauve for slow and precise confirmation in details, particularly focusing on the targeted fragments' gain, lose, and rearrangement, etc..

To conclude, methodologically, the universal genome fingerprint analysis (UGFA) through UGFM-TGCC-SCG (Figure 13) and the pair-wisely genome comparison through Mauve (Figure 14) could be compatible in a manner of sequential operations. In other words, the UGFM-TGCC-SCG method not only could handle exceptional situations for a large set of genomes, but also could facilitate the efficiency of integrating Mauve into performing our so-called systematic comparative genomics, particularly, in terms of in-one-sitting for a set of sequences with farther divergences (chromosome, plasmid, and phage, if applicable) over large size ranges (e.g., 6Kbp ~ 4Mbp). In other words, any component with too farther divergence could be rapidly pre-screened out by UGFM-TGCC-SCG, which could guide on the selection of appropriate subsets of components for subsequent comparison by Mauve.

**Prospective in future for universal genome fingerprint analysis**

*Genome fingerprints and the concepts of strain and species*

"Strain" should be the most fundamental unit for taxonomy. The concise definition of type strain should be crucial for assigning type species, type genus, type family and beyond. Any deep conflicts in arguable strains would eventually shape the assigned species or beyond. Unfortunately, it was so critical but had been overlooked by literatures. To our knowledge to date, no efficient method could provide full description about a type strain, nor was there common agreement upon how to define a species [26, 34, 35]. We anticipated that genomics would be the solid foundation for these issues, as it had re-constructed the concepts of numbers - two or more instead of only one - of chromosomes and replication *ori* points. For example, in the present study, by using our method *GenomeFingerprinter*, we created the whole genome fingerprints (Figure 12, 13) for all eight genetic components in two arguable strains (Table 4) and fundamentally demonstrated that they were not identical (Figure 12, 13) and should probably belong to the same species that needed more characterizations yet. These findings supported the proposal that genome sequence information should be considered in refining an arguable "strain" or "species" in the taxonomy of halophilic Archaea [34]. We agreed with the promotion that in the long run, the definition for a "species" in modern microbiology needed intensive revisions in light of genomics to unify inevitable conflicts in nomenclature system, particularly, in halophilic Archaea [34, 35]. We would further recommend that all genetic components should be included when referring to genomic information for discussing unambiguous taxonomy although to what extent chromosome, plasmid and phage plays roles, respectively, still remained unclear and to be negotiated at current knowledge level [26, 34, 35]. We believed that using UGFM-TGCC-SCG method to concisely resolve the arguments between closely related strains (Figure 13) as well as among farther divergence species or genera (Figure 10, 11, 12, 13) would be one of the crucial steps forwarding to modern microbial nomenclature in the post-genomic era.

## *Type UGFM-TGCC fingerprint for type strain*

We would recommend that any arguable strains should not be judged identical or different only based upon partial information from bulky traditional features such as phenotype and genotype including 16S rRNA, AFLP, PCR-RFLP, ISs, MITEs, etc. [26, 27, 31, 34]. It should also be true for defining a type strain. Theoretically, we would define a type strain or name a new isolate or refine an arguable strain or construct a refined-version for modern microbial taxonomy based upon all unambiguous information from total genome sequences (i.e., chromosome, plasmid, and phage, if applicable). Practically, at least, a type strain should have a meaningful genomic signature. For example, the UGFM-TGCC fingerprint (Figure 10, 11, 12, 13) would be effective to provide a "type strain" with a "type UGFM-TGCC fingerprint" which is simple, standard, and meaningful.

Interestingly, to date, the list of genomes sequenced does not include that of the type strain of *Halobacterium salinarum* (ATCC 33171), the type species of the type genus of the family and the order [36]. It is regretted that no genomic information is available for the nomenclatural type [36]. We expect the community should consider of sequencing more type strains in order to set up a solid foundation for refining modern Archaea taxonomy, which would be invaluable for the next generation of community to understand deeply, research systematically and use efficiently of such amazing bio-resources. Once the "type genome sequence" for the "type strain" is available, the "type UGFM-TGCC fingerprint" can be made by using our methods, as what we did for two related strains *Halobacterium* NRC-1 and *Halobacterium salinarum* R1 (Figure 13) and five diverse strains crossing five genera (Figure 10).

Overall, the family *Halobacteriaceae* consisted of 36 genera with 129 species standing in nomenclature (as of November 2011) [26], but only sixteen strains representing sixteen genera had been sequenced and deposited in GenBank (as of February 2013), including eighteen chromosomes and thirty-six plasmids (Table 2). By using our method of UGFM (I), UGFM-TGCC (II), and UGFM-TGCC-SCG (III), we created the whole genome fingerprints (Figure 10, 11, 12, 13) for all fifty-four genome sequences. Our results provided new insights into critical problematic issues in halophilic Archaea genomics, comparative genomics, and taxonomy [26, 34]. That was a great step on initiatives. We expected more pioneering works to be done. In short, the present paper provided a new method (*GenomeFingerprinter*, Figure 1) for rapidly computing, geometrically visualizing, and intuitively comparing sequences at fingerprint level, and hence established a new method (universal genome fingerprint analysis (UGFA), Figure 6) for systematic comparative genomics, which would be invaluable for the first strategic step forwarding to microbial genomics, comparative genomics, phylogenetics, and taxonomy in the light of post-genomics. We anticipated that our methods could be widely applicable to systematic comparative genomics.

**Conclusions**

We created a new method, *GenomeFingerprinter*, to unambiguously produce three-dimensional coordinates from a sequence, followed by one three-dimensional plot and six two-dimensional trajectory projections to illustrate whole genome fingerprints. We further developed a set of concepts and tools (3D-P, 2D-TP, GF, GFM, P-GFM, S-GFM, UGFM, TGCC, UGFM-TGCC, SCG, and UGFM-TGCC-SCG), and thereby established a new method, universal genome fingerprint analysis (UGFA). We demonstrated their applications through case studies on over a hundred of genome sequences. Particularly, by using UGFM (I), UGFM-TGCC (II), and UGFM-TGCC-SCG (III), we compared a number of genome sequences crossing different genetic category (chromosome, plasmid, and phage; bacterium, archaeal bacterium, and virus) with farther divergences over large size ranges (6Kbp~5Mbp), which we called as systematic comparative genomics, giving new insights into critical problematic issues in microbial genomics. We anticipated that our methods could be widely applicable to systematic comparative genomics in the post-genomic era.

**Materials**

Genome sequences used in this study were downloaded from NCBI or were derived from this study were list in Table 2, 3.

**Methods**

We implemented our method into an in-house script, GenomeFingerprinter.exe. It will be available upon request to the corresponding author. Zplotter (v1.0) and Mauve (v2.3.1) used in this study can be downloaded from links: Zplotter.exe at http://tubic.tju.edu.cn/zcurve/ and Mauve at http://gel.ahabs.wisc.edu/mauve/. To plot graphics from coordinates, any public graphic tool can be used.

**Abbreviations**

bps: base pairs; Kbp: kilo-bps; Mbp: Mega-bps; TB: target base; FB: focusing base; RD: relative distance; WRD: weighted relative distance; SWRD: sum of the weighted relative distances; 3D-P: three-dimensional plot; 2D-TP: two-dimensional trajectory projections; GF: genome fingerprint; GFM: genome fingerprint map; P-GFM: primary genome fingerprint map; S-GFM: secondary genome fingerprint map; UGFM: universal genome fingerprint map; TGCC: total genetic component configuration; UGFM-TGCC: universal genome fingerprint map of total genetic component configuration; SCG: systematic comparative genomics; UGFM-TGCC-SCG: universal genome fingerprint map of total genetic component configuration based systematic comparative genomics; UGFA: universal genome fingerprint analysis

**Acknowledgements**

This work was supported by National High Technology Research & Development Project (2006AA09Z420), National Science and Technology Major Project of China (2011ZX08011005) grants to YA. HA was a recipient of Guangzhou Municipal Science Ambassador Scholarship.

**Authors' contributions**

Conceived and designed the experiments: YA and FM. Performed experiments: LZ (Perl) and HA (Perl and Java) initiated models and scripts; HA reconstructed mathematic models and algorithms, designed, implemented, tested scripts, and constructed system; YA, FM and HA initiated, developed and confined the conceptual frameworks for biological research contents; HA and YA performed computing and collected data. Analyzed the data: HA, YA and FM. Wrote the paper: YA, HA and FM.

**Competing interests**

The author(s) declare that they have no competing interests.

**Figures and Legends**

**Figure 1. A mathematic model for getting coordinates ($x_n$, $y_n$, $z_n$) from a circular genome sequence.** It is arbitrarily starting at the $n^{th}$ base as the chosen target base (TB) and moving to the $m^{th}$ base as a focusing base (FB).

**Figure 2. Three-dimensional plot (3D-P) and primary genome fingerprint map (P-GFM) of fragmental genome sequences of chromosomes in *Escherichia coli* strains.** (A). K-12/W3110 [AC_000091]F7; (B). BL21(DE3)pLysS AG [NC_012947]F1; (C). BL21(DE3)pLysS AG [NC_012947]F5; (D). O55:H7/CB 9615 [NC_013941]F1.

**Figure 3. Six two-dimensional trajectory projections (2D-TP) and secondary genome fingerprint maps (S-GFMs) for *E. coli* K-12/W3110 chromosome [AC_000091].** (A). Projection with $x_n \sim n$; (B). Projection with $y_n \sim n$; (C). Projection with $z_n \sim n$; (D). Projection with $x_n \sim y_n$; (E). Projection with $x_n \sim z_n$; (F). Projection with $y_n \sim z_n$.

**Figure 4. Universal genome fingerprint map (UGFM) for overall comparison of genome fingerprints.** (A). Similar: *Sulfolobus islandicus* M.14.25 [NC_012588] and M.16.4 [NC_012726]; (B). Partly similar: *S. islandicus* Y.N.15.51 [NC_012623] and *Methanococcus voltae* A3 [NC_014222]; (C). Different: *S. islandicus* Y.G.57.14 [NC_012622] and *Methanosphaera stadtmanae* 3091 [NC_007681]; (D). Mixture: (total twelve fragmental sequences (Table 2): 91.1.1, 91.1.61, 91.6.59, 913.1.77, 913.5.57, 4431.1.70, 7946.4.7, 10473.1.74, 10473.4.57, 10498.4.86, 12947.1.50,

13941.2.60.

**Figure 5. Application of universal genome fingerprint map (UGFM) for comparison among a number of genomes in-one-sitting.** The twelve fragmental genome sequences (list in Table 2) were shown in UGFM. The primary genome fingerprint map (P-GFM) of each sequence was classified into different groups solely based upon its location in UGFM: Group (A) (91.1.61, 913.1.77 and 10473.1.74), Group (B) (91.6.59, 913.5.57 and 13941.2.60), Group (C) (7946.4.7 and 12947.1.50), Group (D) (10498.4.86), Group (E) (91.1.1), and Group (F) (4431.1.70). Note that each sequence showed quite different views between its own P-GFM and that in UGFM simply because of the scale-down and view-angle rotation effect in UGFM, which ensured for larger number of objects to be compared in-one-sitting.

**Figure 6. Diagram of conceptual framework for universal genome fingerprint analysis (UGFA).** Our methods consisted of *GenomeFingerprinter* and universal genome fingerprint analysis (UGFA); the former was the fundamental for the latter. The objects could be a single genome sequence, or a number of genome sequences crossing different genetic category (e.g., chromosome, plasmid, phage) in a strain, or a number of sequences of genetic components in a cluster of strains crossing different genetic category (e.g., bacterium, archaeal bacterium, virus). The UGFA method was composed by three subcategories of UGFM (I), UGFM-TGCC (II), and UGFM-TGCC-SCG (III), corresponding to the above three objects, respectively. The core lied in the systematic concepts and tools, which included 3D-P, 2D-TP, P-GFM, S-GFM, UGFM, TGCC, UGFM-TGCC, SCG, and UGFM-TGCC-SCG. Abbreviations: 3D-P: three-dimensional plot; 2D-TP: two-dimensional trajectory projections; GF:

genome fingerprint; GFM: genome fingerprint map; P-GFM: primary genome fingerprint map; S-GFM: secondary genome fingerprint map; UGFM: universal genome fingerprint map; TGCC: total genetic component configuration; UGFM-TGCC: universal genome fingerprint map of total genetic component configuration; SCG: systematic comparative genomics; UGFM-TGCC-SCG: universal genome fingerprint map of total genetic component configuration based systematic comparative genomics; UGFA: universal genome fingerprint analysis

**Figure 7. UGFM (I) of five archaeal strains (each having only one chromosome with size ranging of 2.6 ~4.1 Mbp) crossing five genera of halophilic Archaea.** *Halomonas elongate* DSM 2581 [NC_014532], *Halorhodospira halophilia* SL1 [NC_008789], *Halorhabdus utahensis* DSM 12940 [NC_013158], *Halothermothrix orenii* H 168 [NC_011899] and *Halothiobacillus neapolitanus* c2 [NC_013422] had no close lineages confirming their divergences, at genus level.

**Figure 8. UGFM (I) of forty seven genomes of phages in the family of *Microviridae*.** These forty seven phages were close relatives, but most of them were distinguishable at strain level. They were grouped into two major clusters. Cluster (1) included twenty nine strains (WA5, ID11, WA3, WA2, ID41, NC10, WA6, ID12, NC13, NC2, NC6, ID52, ID8, G4, ID2, WA14, ID18, WA45, ID21, NC28, ID62, NC35, NC29, NC3, alpha3, WA13, phiK, ID32, NC19); Cluster (2) included eighteen strains (NC16, NC5, NC37, ID1, NC7, NC1, NC11, ID22, S13, phiX174, WA11, WA4, ID34, NC41, NC56, WA10, NC51, ID45). The details of phage names were list in Table 3.

**Figure 9. UGFM (I) of twenty four genomes of coronavirus strains.** They were classified into seven clusters. Cluster (1) included the most similar twelve strains of SARS coronavirus ([AY283796], [AY283797], [AY283798], [AY283794], [AY291451], [AY278741], [AY283795], [AY278488], [AY278491], [AY278554], [NC_004718], [AY282752]), tracking with the same UGFM; Cluster (2) included similar four strains of Murine hepatitis virus ([AF201929], [AF208066], [AF208067], [NC_001846]), tracking with the similar UGFM; Cluster (3) was a distinctive Porcine epidemic diarrhea virus strain ([NC_003436]); Cluster (4) was a distinctive Avian infectious bronchitis virus strain ([NC_001451]); Cluster (5) was a distinctive Feline infectious peritonitis virus strain ([NC_002306]); Cluster (6) was a distinctive Human coronavirus strain ([NC_002645]); Cluster (7) included four similar strains of Bovine coronavirus ([AF220295], [u00735], [AF391542], [NC_003045]), tracking with the similar UGFM. These seven clusters were perfectly matched to their biological identity groups (list in Table 3).

**Figure 10. UGFM-TGCC (II) of five archaeal strains crossing four genera of halophilic Archaea.** (A) enlarged vision of those five plasmids and (B) *Halogeometricum boringquense* DSM 11551 (one chromosome [NC_014729] and five plasmids pHBOR02 [NC_014731], pHBOR04 [NC_014732], pHBOR01 [NC_014735], pHBOR03 [NC_014736], pHBOR05 [NC_014737]); (C) enlarged vision of those six plasmids and (D) *Haloterrigena turkmenica* DSM 5511 (one chromosome [NC_013743] and six plasmids pHTUR01 [NC_013744], pHTUR02 [NC_013745], pHTUR03 [NC_013746], pHTUR04 [NC_013747], pHTUR05 [NC_013748], pHTUR06 [NC_013749]); (E) enlarged vision of those three plasmids and (F) *Natrialba magadii* ATCC 43099 (one chromosome [NC_013922] and three plasmids

pNMAG01 [NC_013923], pNMAG02 [NC_013924], pNMAG03 [NC_013925]; (G) *Natrinema pellirubrum* DSM 15624 (one chromosome [NC_019962] and two plasmids pNATPE02 [NC_019963], pNATPE01 [NC_019967]; (H) *Haloquadratum walsbyi* DSM 16790 (one chromosome [NC_008212] and one plasmid PL47 [NC_008213]. These four strains crossing four genera (Table 2) had quite different UGFM-TGCC visions demonstrating their farther lineages at genus level.

**Figure 11. UGFM-TGCC-SCG** (III) **of four archaeal strains (each having multiple chromosomes and plasmids) crossing four genera of halophilic Archaea.** There were two sets of in-one-sitting comparison. One set (A-B) : *Halorubrum lacusprofundii* ATCC49239 (chromosome I [NC_012029], chromosome II [NC_012028], plasmid pHLAC01 [NC_012030]) *vs. Haloarcula marismortui* ATCC43049 (chromosome I [NC_006396], chromosome II [NC_006397], and seven plasmids pNG100 [NC_006389], pNG200 [NC_006390], pNG300 [NC_006391], pNG400 [NC_006392], pNG500 [NC_006393], pNG600 [NC_006394], pNG700 [NC_006395]) focusing on plasmids (A) and a universal system (B); Another set (C-D): *Haloferax vocanii* DS2 [chromosome [NC_013967], plasmid pHV3 [NC_013964], pHV2 [NC_013965], pHV4 [NC_013966], pHV1 [NC_013968] ) *vs. Halomicrobium mukohataei* DSM 12286 (chromosome [NC_013202], plasmid pHmuk01[NC_013201] ) focusing on plasmids (C) and a universal system (D).

**Figure 12. Application of universal genome fingerprint analysis (UGFA) for comparative genomics between two chromosomes of *Halobacterium* sp. NRC-1 [NC_002607] and *Halobacterium salinarum* R1 [NC_010364].** Two arguable strains were compared by using three-dimensional plots ($x_n$~$y_n$~$z_n$) (P-GFM) (A) and

two-dimensional trajectory projections (S-GFM) with different combinations of coordinates: (B) $x_n$~$y_n$; (C) $x_n$~$z_n$; (D) $y_n$~$z_n$; (E) $x_n$~n; (F) $y_n$~n; (G) $z_n$~n; (H) $x_n$~n and $y_n$~n together. Note two arrows showed replication *ori* points, *ori*C1 and *ori* C2; other arrows indicated genome-wide evolution events.

**Figure 13. Application of universal genome fingerprint analysis (UGFA) for systematic comparative genomics (SCG).** The universal genome fingerprint map (UGFM) of total genetic component configurations (TGCC) (UGFM-TGCC) was applied to the systematic comparative genomics (SCG) in-one-sitting. Two strains *Halobacterium* sp. NRC-1 and *Halobacterium salinarum* R1 were compared as a universal system. (A). UGFM-TGCC-SCG for total six plasmids (*Halobacterium* sp. NRC-1 pNRC100 [NC_00001869] and pNRC200 [NC_002608]; *Halobacterium salinarum* R1 PHS1 [NC_010366], PHS2 [NC_010369], PHS3 [NC_010368], and PHS4 [NC_010367]); (B). UGFM-TGCC-SCG for those total six plasmids and two chromosomes (*Halobacterium* sp. NRC-1 [NC_002607] and *Halobacterium salinarum* R1 [NC_010364]). Note, even within the same strain, chromosome and plasmid showed distinctive lineage divergences. In other words, there was no correlation between chromosome and plasmid within a certain strain, i.e., without any binding to a certain strain, indicating possible independent evolution among chromosomes and plasmids.

**Figure 14. Mauve snapshots for pair-wisely genome comparisons between two strains *Halobacterium* sp. NRC-1 and *Halobacterium salinarum* R1 considering of TGCC as a universal system.** progressiveMauve mode analysis could compare in-one-sitting all eight components of NRC-1 and R1 strains showing bare

homologous (A). Mauve mode analysis failed because it stopped alignment due to no homologous; but it worked well, separately, with (B) six plasmids (the inner window) (*Halobacterium* sp. NRC-1 pNRC100 [NC_00001869] and pNRC200 [NC_002608]; *Halobacterium salinarum* R1 PHS1 [NC_010366], PHS2 [NC_010369], PHS3 [NC_010368], and PHS4 [NC_010367]), and (C) two chromosomes (*Halobacterium* sp. NRC-1 [NC_002607] and *Halobacterium salinarum* R1 [NC_010364]). Mauve mode analysis could clearly reveal the relationships among six plasmids and between two chromosomes, separately.

**Figure 15. Comparisons between two chromosomes of *Halobacterium* sp. NRC-1 [NC_002607] and its derivative form (NC_002607_RC) with different cutting-point.** (A). Comparison via *GenomeFingerprinter*; (B). Comparison via Zplotter with $z_n$'; (C). Comparison via Zplotter with $z_n$.

# Tables and Captions

## Table 1. Coordinates of an artificial sample sequence

| # bp / point | $x_n$ | $y_n$ | $z_n$ |
|---|---|---|---|
| 1 | 0.7 | 8.1 | -4.1 |
| 2 | -0.3 | 8.7 | -4.9 |
| 3 | 0.7 | 9.3 | -3.7 |
| 4 | -0.3 | 9.9 | -4.5 |
| 5 | -1.3 | 8.5 | -3.3 |
| 6 | -0.3 | 7.1 | -4.1 |
| 7 | 0.7 | 7.7 | -2.9 |
| 8 | -0.3 | 8.3 | -3.7 |
| 9 | 0.7 | 6.9 | -4.5 |
| 10 | -0.3 | 7.5 | -5.3 |
| 11 | 0.7 | 8.1 | -4.1 |
| 12 | -0.3 | 8.7 | -4.9 |
| 13 | 0.7 | 9.3 | -3.7 |
| 14 | -0.3 | 9.9 | -4.5 |
| 15 | -1.3 | 8.5 | -3.3 |
| 16 | -0.3 | 7.1 | -4.1 |
| 17 | 0.7 | 7.7 | -2.9 |
| 18 | -0.3 | 8.3 | -3.7 |
| 19 | 0.7 | 6.9 | -4.5 |
| 20 | -0.3 | 7.5 | -5.3 |
| 21 | 0.7 | 8.1 | -4.1 |
| 22 | -0.3 | 8.7 | -4.9 |
| 23 | 0.7 | 9.3 | -3.7 |
| 24 | -0.3 | 9.9 | -4.5 |
| 25 | -1.3 | 8.5 | -3.3 |
| 26 | -0.3 | 7.1 | -4.1 |
| 27 | 0.7 | 7.7 | -2.9 |
| 28 | -0.3 | 8.3 | -3.7 |
| 29 | 0.7 | 6.9 | -4.5 |
| 30 | -0.3 | 7.5 | -5.3 |
| 31 | 0.7 | 8.1 | -4.1 |
| 32 | -0.3 | 8.7 | -4.9 |
| 33 | 0.7 | 9.3 | -3.7 |
| 34 | -0.3 | 9.9 | -4.5 |
| 35 | -1.3 | 8.5 | -3.3 |
| 36 | -0.3 | 7.1 | -4.1 |
| 37 | 0.7 | 7.7 | -2.9 |
| 38 | -0.3 | 8.3 | -3.7 |
| 39 | 0.7 | 6.9 | -4.5 |
| 40 | -0.3 | 7.5 | -5.3 |

## Table 2. Features of genome sequences from bacteria and archaeal bacteria used in this study

| Species and Strain | Sequence ID | Type | Size (bps) |
|---|---|---|---|
| **Downloaded from FTP.ncbi.nlm.nih.gov [GenBank]** | | | |
| *Escherichia coli* K-12/W3110 | AC_000091 NC_007779 | Chromosome* | 4646332 |
| *Escherichia coli* K-12/DH10B | NC_010473 | Chromosome* | 4686137 |
| *Escherichia coli* K-12/MG1655 | NC_000913 | Chromosome* | 4639675 |
| *Escherichia coli* BL21 (DE3) pLysSAG | NC_012947 | Chromosome* | 4570938 |
| *Escherichia coli* O55:H7/CB9615 | NC_013941 | Chromosome* | 5386352 |
| *Escherichia coli* UTI89 | NC_007946 | Chromosome* | 5065741 |
| *Escherichia coli* CFT073 | NC_004431 | Chromosome* | 5231428 |
| *Escherichia coli* SMS-3-5 | NC_010498 | Chromosome* | 5068389 |
| *Sulfolobus islandicus* M.14.25 | NC_012588 | Chromosome* | 2608832 |
| *Sulfolobus islandicus* M.16.4 | NC_012726 | Chromosome* | 2586647 |
| *Sulfolobus islandicus* Y.N.15.51 | NC_012623 | Chromosome* | 2812165 |
| *Sulfolobus islandicus* Y.G.57.14 | NC_012622 | Chromosome* | 2702058 |
| *Methanococcus voltae* A3 | NC_014222 | Chromosome* | 1936387 |
| *Methanosphaera stadtmanae* DSM 3091 | NC_007681 | Chromosome* | 1767403 |
| *Halomonas elongate* DSM 2581 | NC_014532 | Chromosome*[a] | 4119315 |
| *Halorhodospira halophilia* SL1 | NC_008789 | Chromosome*[a] | 2716716 |
| *Halorhabdus utahensis* DSM 12940 | NC_013158 | Chromosome*[a] | 3161321 |
| *Halothermothrix orenii* H 168 | NC_011899 | Chromosome*[a] | 2614977 |
| *Halothiobacillus neapolitanus* c2 | NC_013422 | Chromosome*[a] | 2619785 |
| *Halogeometricum boringquense* DSM 11551 | NC_014729 | Chromosome*[b] | 2860838 |
| *Halogeometricum boringquense* DSM 11551 | NC_014731 | plasmid pHBOR02[b] | 343853 |
| *Halogeometricum boringquense* DSM 11551 | NC_014732 | plasmid pHBOR04[b] | 197618 |
| *Halogeometricum boringquense* DSM 11551 | NC_014735 | plasmid pHBOR01[b] | 367369 |
| *Halogeometricum boringquense* DSM 11551 | NC_014736 | plasmid pHBOR03[b] | 213355 |
| *Halogeometricum boringquense* DSM 11551 | NC_014737 | plasmid pHBOR05[b] | 17786 |
| *Haloterrigena turkmenica* DSM 5511 | NC_013743 | Chromosome*[b] | 3944596 |
| *Haloterrigena turkmenica* DSM 5511 | NC_013744 | plasmid pHTUR01[b] | 708474 |
| *Haloterrigena turkmenica* DSM 5511 | NC_013745 | plasmid pHTUR02[b] | 419558 |
| *Haloterrigena turkmenica* DSM 5511 | NC_013746 | plasmid pHTUR03[b] | 183364 |
| *Haloterrigena turkmenica* DSM 5511 | NC_013747 | plasmid pHTUR04[b] | 174400 |
| *Haloterrigena turkmenica* DSM 5511 | NC_013748 | plasmid pHTUR05[b] | 72078 |
| *Haloterrigena turkmenica* DSM 5511 | NC_013749 | plasmid pHTUR06[b] | 16041 |
| *Natrialba magadii* ATCC 43099 | NC_013922 | Chromosome*[b] | 3805456 |
| *Natrialba magadii* ATCC 43099 | NC_013923 | plasmid pNMAG01[b] | 383753 |
| *Natrialba magadii* ATCC 43099 | NC_013924 | plasmid pNMAG02[b] | 258593 |
| *Natrialba magadii* ATCC 43099 | NC_013925 | plasmid pNMAG03[b] | 59323 |
| *Natrinema pellirubrum* DSM 15624 | NC_019962 | Chromosome*[b] | 3844629 |
| *Natrinema pellirubrum* DSM 15624 | NC_019963 | plasmid pNATPE02[b] | 279762 |
| *Natrinema pellirubrum* DSM 15624 | NC_019967 | plasmid pNATPE01[b] | 291912 |
| *Haloquadratum walsbyi* DSM 16790 | NC_008212 | Chromosome*[b] | 3177244 |
| *Haloquadratum walsbyi* DSM 16790 | NC_008213 | plasmid PL47[b] | 47537 |
| *Halorubrum lacusprofundii* ATCC49239 | NC_012029 | Chromosome I*[c] | 2774371 |
| *Halorubrum lacusprofundii* ATCC49239 | NC_012028 | Chromosome II*[c] | 533457 |
| *Halorubrum lacusprofundii* ATCC49239 | NC_012030 | plasmid pHLAC01[c] | 437500 |
| *Haloarcula marismortui* ATCC43049 | NC_006396 | Chromosome I*[c] | 3176463 |
| *Haloarcula marismortui* ATCC43049 | NC_006397 | Chromosome II*[c] | 292165 |
| *Haloarcula marismortui* ATCC43049 | NC_006389 | plasmid pNG100[c] | 33779 |
| *Haloarcula marismortui* ATCC43049 | NC_006390 | plasmid pNG200[c] | 33930 |
| *Haloarcula marismortui* ATCC43049 | NC_006391 | plasmid pNG300[c] | 40086 |
| *Haloarcula marismortui* ATCC43049 | NC_006392 | plasmid pNG400[c] | 50776 |
| *Haloarcula marismortui* ATCC43049 | NC_006393 | plasmid pNG500[c] | 134574 |
| *Haloarcula marismortui* ATCC43049 | NC_006394 | plasmid pNG600[c] | 157549 |
| *Haloarcula marismortui* ATCC43049 | NC_006395 | plasmid pNG700[c] | 416420 |
| *Halomicrobium mukohataei* DSM 12286 | NC_013202 | Chromosome*[c] | 3154923 |
| *Halomicrobium mukohataei* DSM 12286 | NC_013201 | plasmid pHmuk01[c] | 225032 |
| *Haloferax vocanii* DS2 | NC_013967 | Chromosome*[c] | 2888440 |
| *Haloferax vocanii* DS2 | NC_013964 | plasmid pHV3[c] | 444162 |
| *Haloferax vocanii* DS2 | NC_013965 | plasmid pHV2[c] | 6450 |
| *Haloferax vocanii* DS2 | NC_013966 | plasmid pHV4[c] | 644869 |
| *Haloferax vocanii* DS2 | NC_013968 | plasmid pHV1[c] | 86308 |
| *Halobacterium* sp.NRC-1 | NC_002607 | Chromosome*[d] | 2014239 |
| *Halobacterium* sp.NRC-1 | NC_001869 | Plasmid pNRC100[d] | 191346 |
| *Halobacterium* sp.NRC-1 | NC_002608 | Plasmid pNRC200[d] | 365425 |
| *Halobacterium salinarum* R1 | NC_010364 | Chromosome*[d] | 2000962 |
| *Halobacterium salinarum* R1 | NC_010366 | Plasmid PHS1[d] | 147625 |
| *Halobacterium salinarum* R1 | NC_010369 | Plasmid PHS2[d] | 194963 |
| *Halobacterium salinarum* R1 | NC_010368 | Plasmid PHS3[d] | 284332 |
| *Halobacterium salinarum* R1 | NC_010367 | Plasmid PHS4[d] | 40894 |
| **Derivatives created in this study [based on those sequences from GenBank]** | | | |
| *Escherichia coli* K-12/W3110-91.1.1 | 91.1.1 | Chromosome fragment | 227694 |
| *Escherichia coli* K-12/W3110-91.1.61 | 91.1.61 | Chromosome fragment | 324260 |
| *Escherichia coli* K-12/W3110-91.6.59 | 91.6.59 | Chromosome fragment | 410186 |
| *Escherichia coli* K-12/W3110-91.F7 | 91.7 | Chromosome fragment | 953958 |
| *Escherichia coli* K-12/MG1655-913.1.77 | 913.1.77 | Chromosome fragment | 331163 |
| *Escherichia coli* K-12/MG1655-913.5.57 | 913.5.57 | Chromosome fragment | 408963 |
| *Escherichia coli* CFT073-4431.1.70 | 4431.1.70 | Chromosome fragment | 401260 |
| *Escherichia coli* UTI89-7946.4.7 | 7946.4.7 | Chromosome fragment | 518065 |
| *Escherichia coli* K-12/DH10B -10473.1.74 | 10473.1.74 | Chromosome fragment | 325622 |
| *Escherichia coli* K-12/DH10B -10473.4.57 | 10473.4.57 | Chromosome fragment | 412818 |
| *Escherichia coli* SMS-3-5-10498.4.86 | 10498.4.86 | Chromosome fragment | 331536 |
| *Escherichia coli* BL21 (DE3) pLysSAG-12947.F1 | 12947.1 | Chromosome fragment | 1759795 |
| *Escherichia coli* BL21 (DE3) pLysSAG-12947.1.50 | 12947.1.50 | Chromosome fragment | 470050 |
| *Escherichia coli* BL21 (DE3) pLysSAG-12947.F5 | 12947.5 | Chromosome fragment | 43254 |
| *Escherichia coli* O55:H7/CB9615-13941.F1 | 13941.1 | Chromosome fragment | 1915479 |
| *Escherichia coli* O55:H7/CB9615-13941.2.60 | 13941.2.60 | Chromosome fragment | 267039 |

\* 32 chromosomes used for calculations as list in Table 5.

[a]UGFM (I): five strains and five genomes (Figure 7)

[b]UGFM-TGCC (II): five strains and twenty two genomes (Figure 10)

[c]UGFM-TGCC-SCG (III): four strains and nineteen genomes (Figure 11)

[d]Case studies: two strains and eight genomes (Figure 12, 13, 14)

**Table 3. Features of genome sequences from viruses and phages used in this study**

| Species and Strain | Sequence ID | Type | Size (bps) |
|---|---|---|---|
| **Downloaded from FTP.ncbi.nlm.nih.gov [GenBank]** | | | |
| WA5: Coliphage WA5 | NC_007847 | Phage chromosome[a] | 5737 |
| ID11: Coliphage ID11 | NC_006954 | Phage chromosome[a] | 5737 |
| WA3: Coliphage WA3 | NC_007845 | Phage chromosome[a] | 5700 |
| WA2: Coliphage WA2 | NC_007844 | Phage chromosome[a] | 5700 |
| ID41: Coliphage ID41 | NC_007851 | Phage chromosome[a] | 5737 |
| NC10: Coliphage NC10 | NC_007854 | Phage chromosome[a] | 5687 |
| WA6: Coliphage WA6 | NC_007852 | Phage chromosome[a] | 5687 |
| ID12: Coliphage ID12 | NC_007853 | Phage chromosome[a] | 5687 |
| NC13: Coliphage NC13 | NC_007849 | Phage chromosome[a] | 5737 |
| NC2: Coliphage NC2 | NC_007848 | Phage chromosome[a] | 5737 |
| NC6: Coliphage NC6 | NC_007855 | Phage chromosome[a] | 5687 |
| ID52: Coliphage ID52 | NC_007825 | Phage chromosome[a] | 5698 |
| ID8: Coliphage ID8 | NC_007846 | Phage chromosome[a] | 5700 |
| G4: Enterobacteria phage G4 | NC_001420 | Phage chromosome[a] | 5737 |
| ID2: Coliphage ID2 | NC_007817 | Phage chromosome[a] | 5644 |
| WA14: Coliphage WA14 | NC_007857 | Phage chromosome[a] | 5644 |
| ID18: Coliphage ID18 | NC_007856 | Phage chromosome[a] | 5644 |
| WA45: Coliphage WA45 | NC_007822 | Phage chromosome[a] | 6242 |
| ID21: Coliphage ID21 | NC_007818 | Phage chromosome[a] | 6242 |
| NC28: Coliphage NC28 | NC_007823 | Phage chromosome[a] | 6239 |
| ID62: Coliphage ID62 | NC_007824 | Phage chromosome[a] | 6225 |
| NC35: Coliphage NC35 | NC_007820 | Phage chromosome[a] | 6213 |
| NC29: Coliphage NC29 | NC_007827 | Phage chromosome[a] | 6439 |
| NC3: Coliphage NC3 | NC_007826 | Phage chromosome[a] | 6273 |
| alpha3: Enterobacteria phage alpha3 | DQ085810 | Phage chromosome[a] | 6177 |
| WA13: Coliphage WA13 | NC_007821 | Phage chromosome[a] | 6242 |
| phiK: Coliphage phiK | NC_001730 | Phage chromosome[a] | 6263 |
| ID32: Coliphage ID32 | NC_007819 | Phage chromosome[a] | 6245 |
| NC19: Coliphage NC19 | NC_007850 | Phage chromosome[a] | 5737 |
| NC16: Coliphage NC16 | NC_007836 | Phage chromosome[a] | 5540 |
| NC5: Coliphage NC5 | NC_007833 | Phage chromosome[a] | 5540 |
| NC37: Coliphage NC37 | NC_007837 | Phage chromosome[a] | 5540 |
| ID1: Coliphage ID1 | NC_007828 | Phage chromosome[a] | 5540 |
| NC7: Coliphage NC7 | NC_007834 | Phage chromosome[a] | 5540 |
| NC1: Coliphage NC1 | NC_007832 | Phage chromosome[a] | 5540 |
| NC11: Coliphage NC11 | NC_007835 | Phage chromosome[a] | 5540 |
| ID22: Coliphage ID22 | NC_007829 | Phage chromosome[a] | 5540 |
| S13: Enterobacteria phage S13 | NC_001424 | Phage chromosome[a] | 5540 |
| phiX174: Coliphage phiX174 | NC_001422 | Phage chromosome[a] | 5540 |
| WA11: Coliphage WA11 | NC_007843 | Phage chromosome[a] | 5541 |
| WA4: Coliphage WA4 | NC_007841 | Phage chromosome[a] | 5540 |
| ID34: Coliphage ID34 | NC_007830 | Phage chromosome[a] | 5540 |
| NC41: Coliphage NC41 | NC_007838 | Phage chromosome[a] | 5540 |
| NC56: Coliphage NC56 | NC_007840 | Phage chromosome[a] | 5540 |
| WA10: Coliphage WA10 | NC_007842 | Phage chromosome[a] | 5540 |
| NC51: Coliphage NC51 | NC_007839 | Phage chromosome[a] | 5540 |
| ID45: Coliphage ID45 | NC_007831 | Phage chromosome[a] | 5540 |
| *SARS coronavirus* TW1 | AY283796 | Virus chromosome[b] | 30137 |
| *SARS coronavirus* Sin2679 | AY283797 | Virus chromosome[b] | 30132 |
| *SARS coronavirus* Sin2748 | AY283798 | Virus chromosome[b] | 30137 |
| *SARS coronavirus* Sin2774 | AY283794 | Virus chromosome[b] | 30137 |
| *SARS coronavirus* Sin2500 | AY291451 | Virus chromosome[b] | 30155 |
| *SARS coronavirus* Urbani | AY278741 | Virus chromosome[b] | 30153 |
| *SARS coronavirus* Sin2677 | AY283795 | Virus chromosome[b] | 30131 |
| *SARS coronavirus* BJ01 | AY278488 | Virus chromosome[b] | 30151 |
| *SARS coronavirus* HKU-39849 | AY278491 | Virus chromosome[b] | 30168 |
| *SARS coronavirus* CUHK-W1 | AY278554 | Virus chromosome[b] | 30162 |
| *SARS coronavirus* | NC_004718 | Virus chromosome[b] | 30178 |
| *SARS coronavirus* CUHK-Su10 | AY282752 | Virus chromosome[b] | 30162 |
| *Murine hepatitis virus* strain 2 | AF201929 | Virus chromosome[b] | 31724 |
| *Murine hepatitis virus* strain Penn 97-1 | AF208066 | Virus chromosome[b] | 31558 |
| *Murine hepatitis virus* strain ML-10 | AF208067 | Virus chromosome[b] | 31681 |
| *Murine hepatitis virus* strain A59 | NC_001846 | Virus chromosome[b] | 31806 |
| *Porcine epidemic diarrhea virus* | NC_003436 | Virus chromosome[b] | 28435 |
| *Avian infectious bronchitis virus* | NC_001451 | Virus chromosome[b] | 28004 |
| *Feline infectious peritonitis virus* | NC_002306 | Virus chromosome[b] | 29776 |
| *Human coronavirus* 229E | NC_002645 | Virus chromosome[b] | 27709 |
| *Bovine coronavirus* strain Quebec | AF220295 | Virus chromosome[b] | 31546 |
| *Bovine coronavirus* strain Mebus | u00735 | Virus chromosome[b] | 31477 |
| *Bovine coronavirus* isolate BCoV-LUN | AF391542 | Virus chromosome[b] | 31473 |
| *Bovine coronavirus* | NC_003045 | Virus chromosome[b] | 31473 |

[a]UGFM (I): forty seven strains and forty seven genomes in phage (Figure 8)

[b]UGFM (I): twenty four strains and twenty four genomes in virus (Figure 9)

**Table 4. Features of total genetic component configurations of strains NRC-1 and R1**

| Components | NRC-1 | R1 |
|---|---:|---:|
| Chromosome (bp) | NC_002607 | NC_010364 |
|  | 2014239 | 2000962 |
| Plasmid 1 (bp) | pNRC100 | PHS1 |
|  | NC_001869 | NC_010366 |
|  | 191346 | 147625 |
| Plasmid 2 (bp) | pNRC200 | PHS2 |
|  | NC_002608 | NC_010369 |
|  | 365425 | 194963 |
| Plasmid 3 (bp) |  | PHS3 |
|  | - | NC_010368 |
|  |  | 284332 |
| Plasmid 4 (bp) |  | PHS4 |
|  | - | NC_0103667 |
|  |  | 40,894 |
| Total (bp) | 2571010 | 2668776 |

**Table 5. *GenomeFingerprinter vs.* Mauve**

| Number | *GenomeFingerprinter* | | Mauve |
|---|---|---|---|
| 1 chr. | 1 min cal. | 1 min plot | no valid |
| 2 chr. | 2 min cal. | 2 min plot | 2 min |
| 4 chr. | 4 min cal. | 4 min plot | 8 min |
| 8 chr. | 8 min cal. | 8 min plot | 44 min |
| 16 chr. | 16 min cal. | 16 min plot | 332 min |
| 32 chr. | 32 min cal. | 32 min plot | MO |

* Notes:

1) MO: memory overflow;

2) Samples: 32 chromosomes (chr.) as list (*) in Table 2;

3) Conditions: HP Proliant server DL580-G5 with 16 CPU/8Gb memory.

$n^{th}$

$(n=1, 2, ..., N)$

3'

n

5'

m

unprocessed $m^{th}$: $N-m$

$m^{th}$

$(m=1, 2, ..., N)$

processed $m^{th}$: $m$

Figure
Click here to download high resolution image

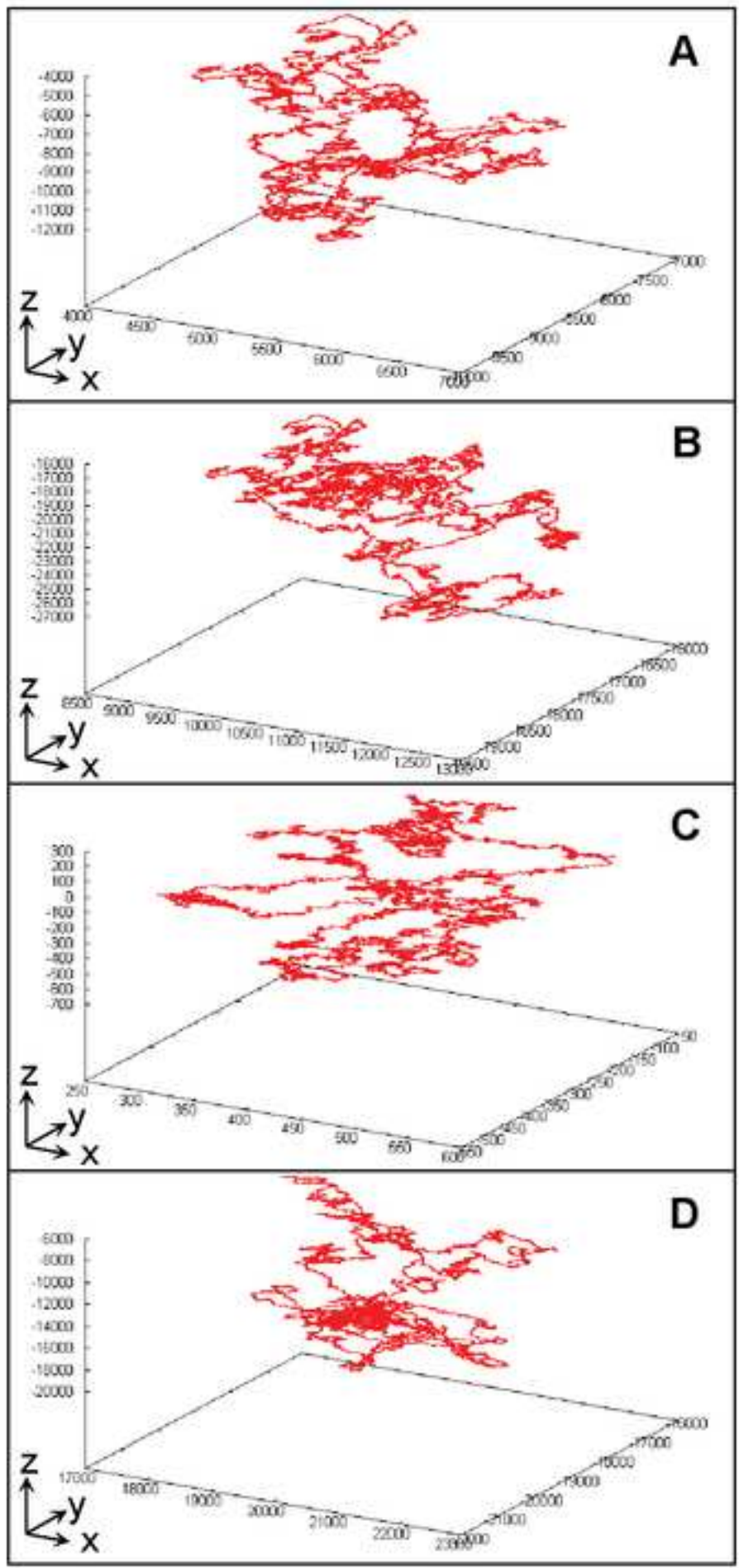

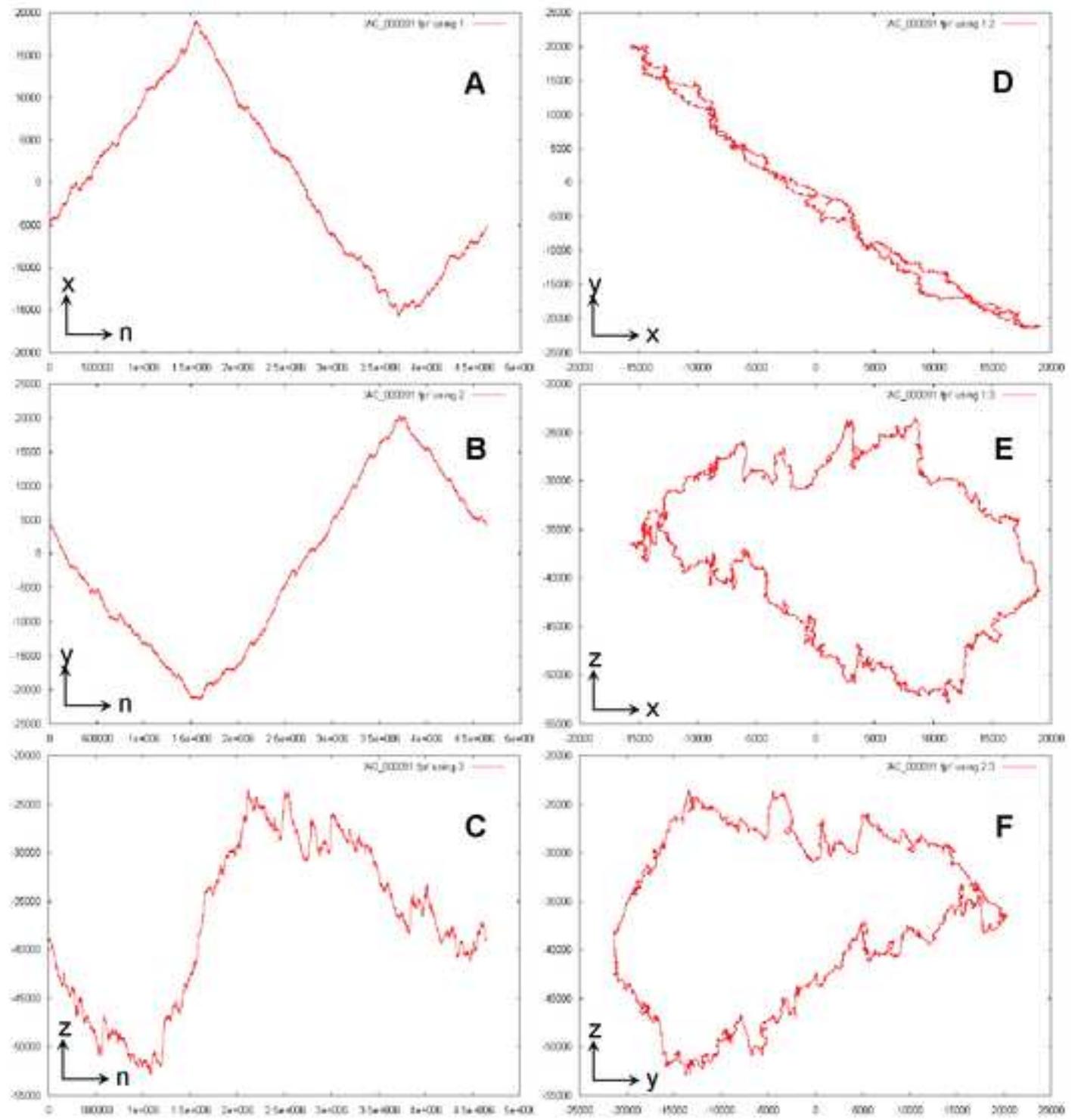

Figure
Click here to download high resolution image

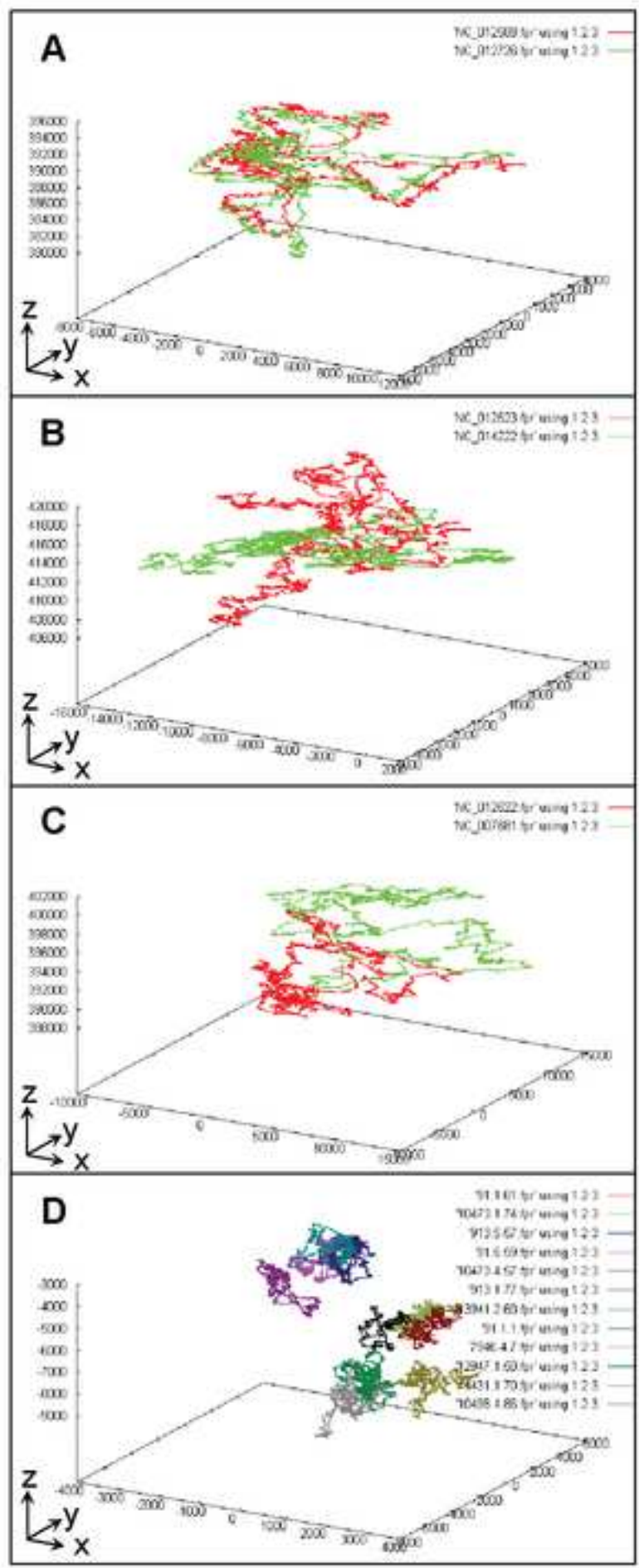

Figure
Click here to download high resolution image

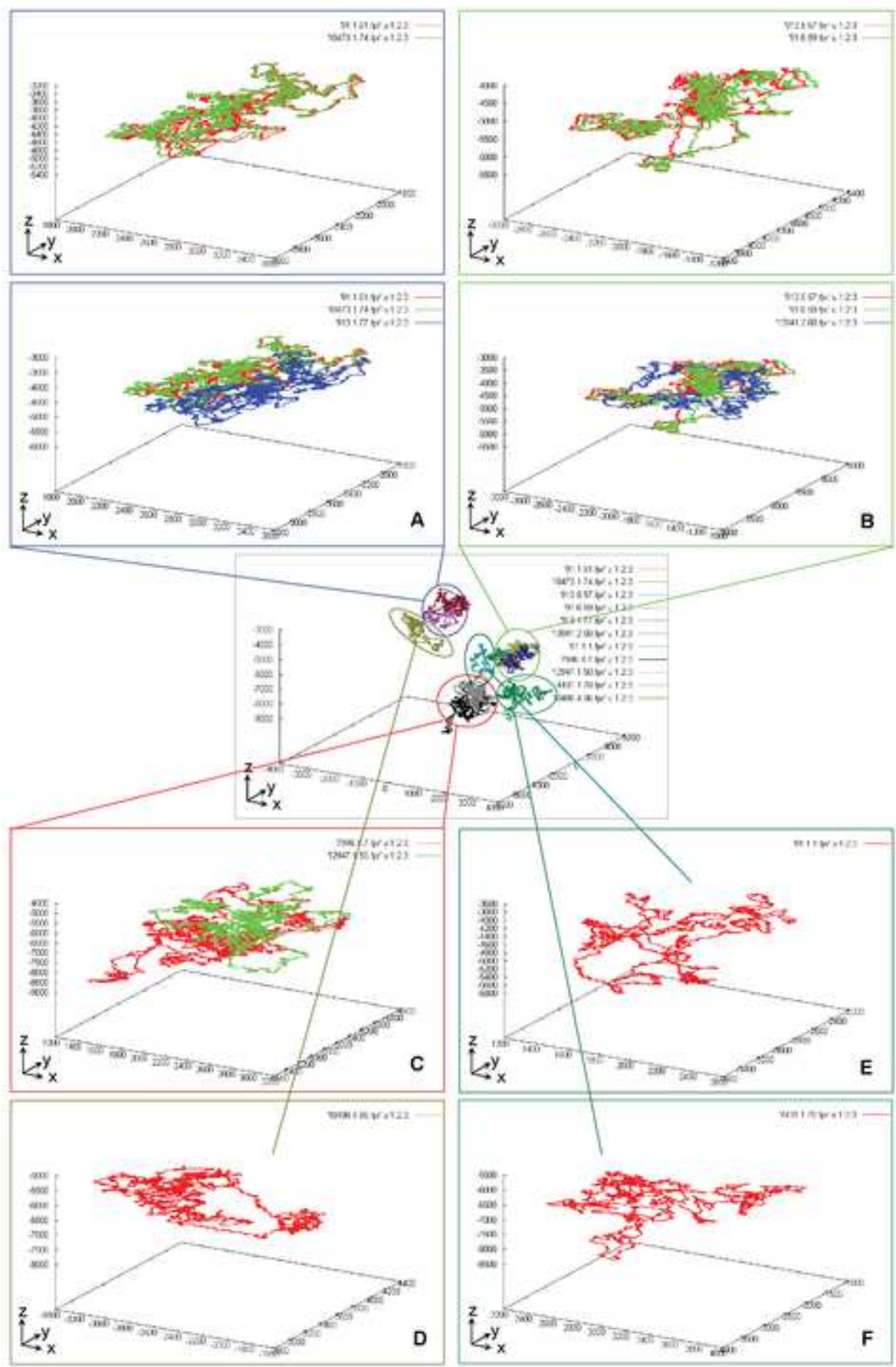

Figure
Click here to download high resolution image

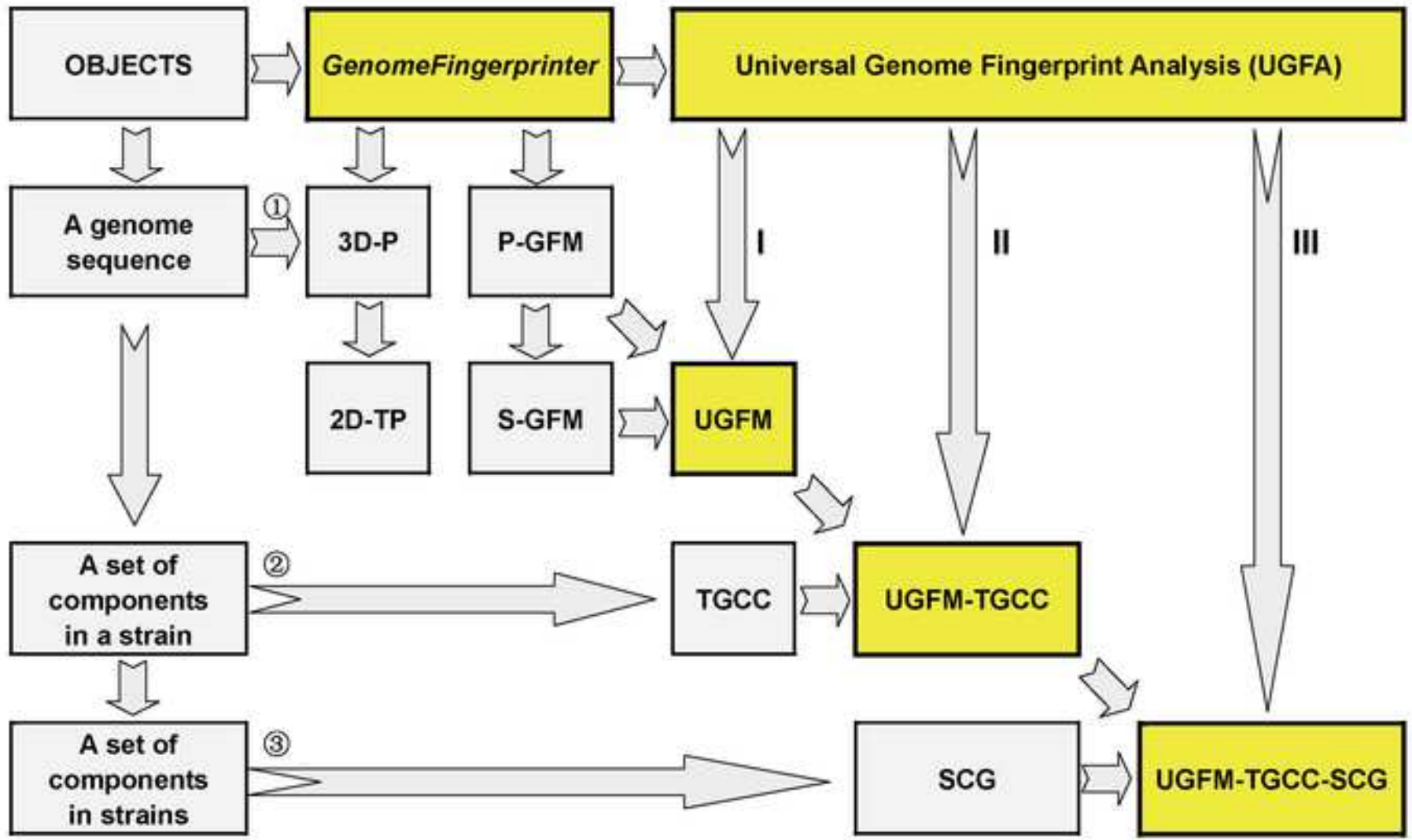

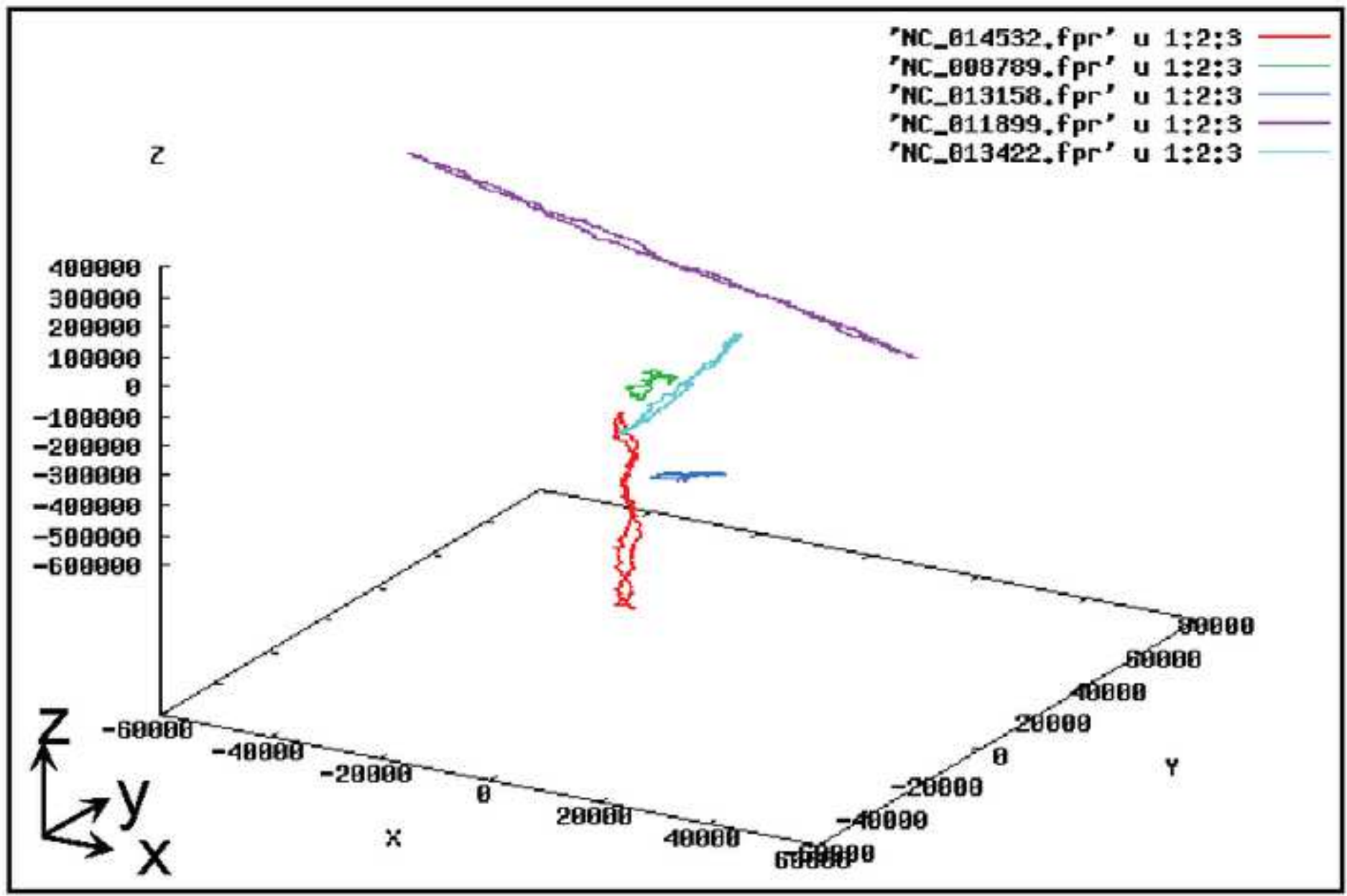

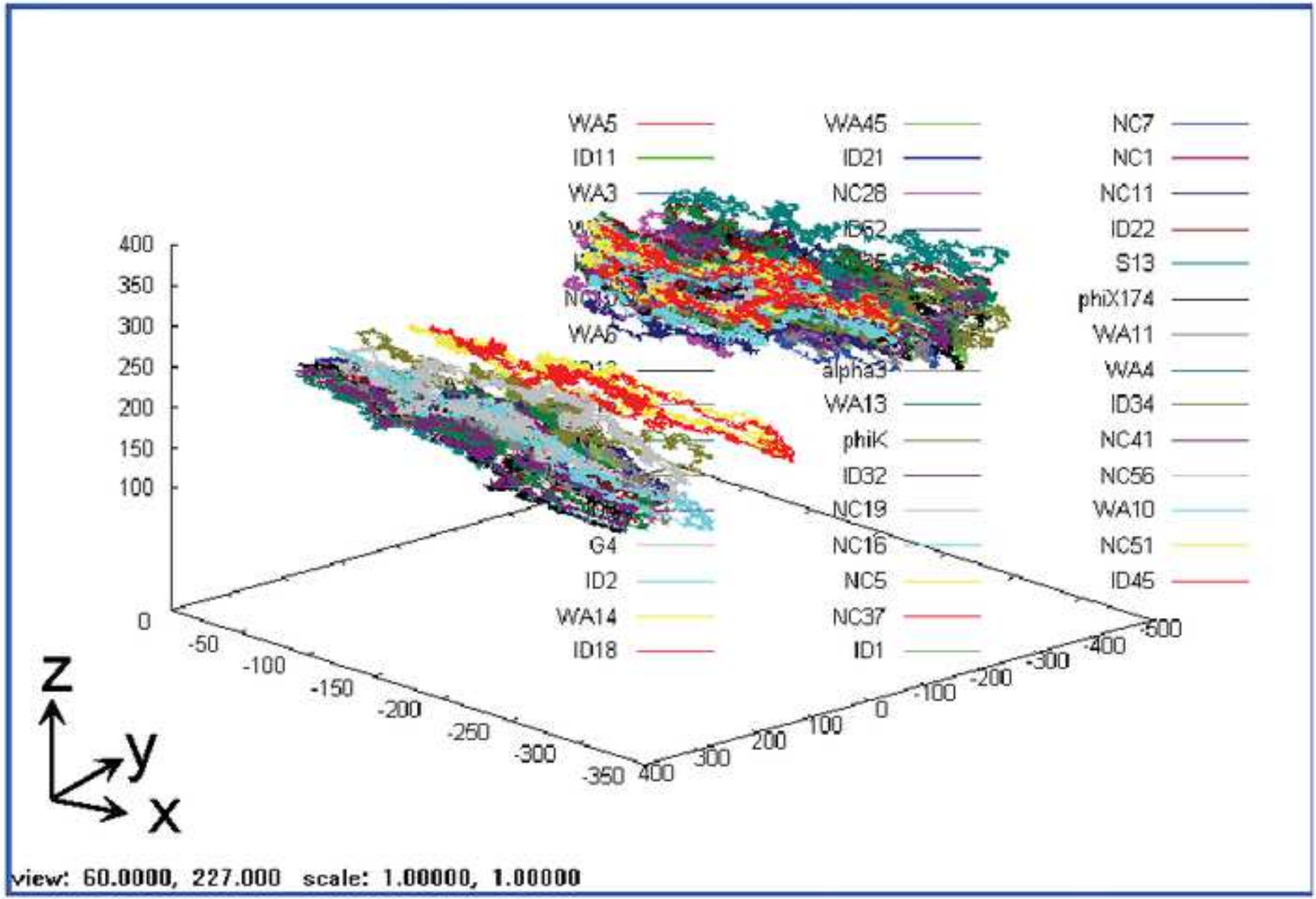

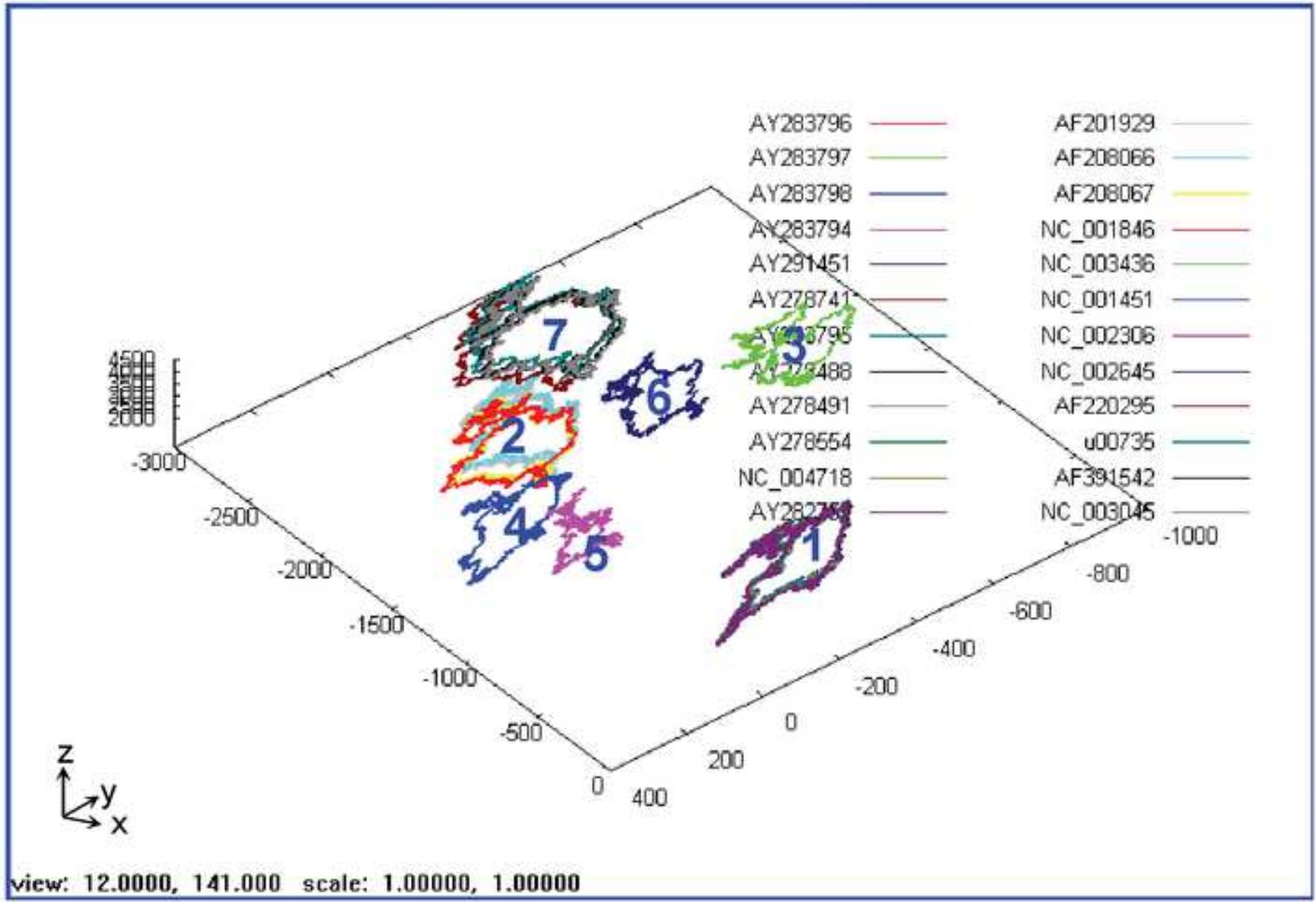

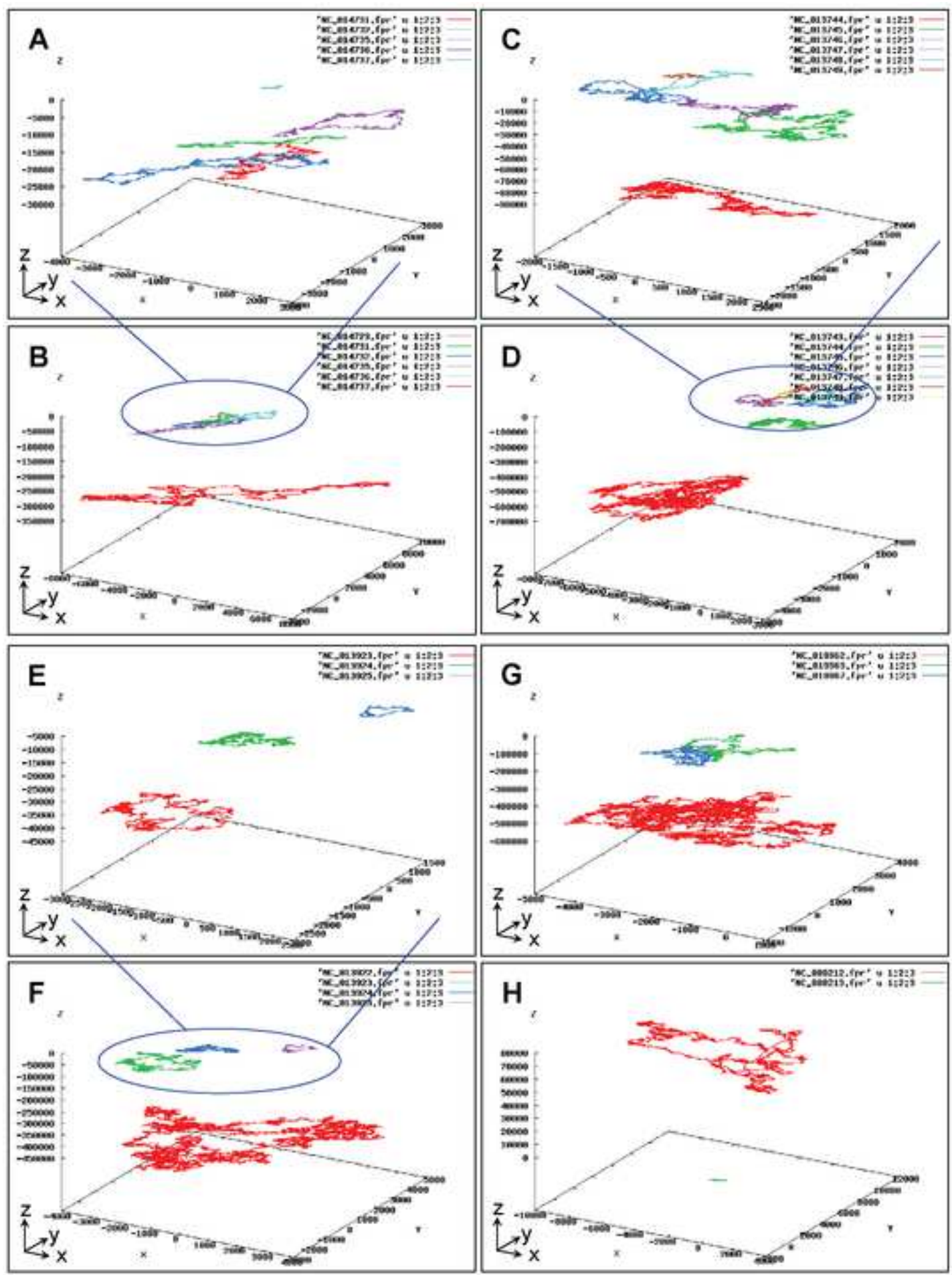

Figure
Click here to download high resolution image

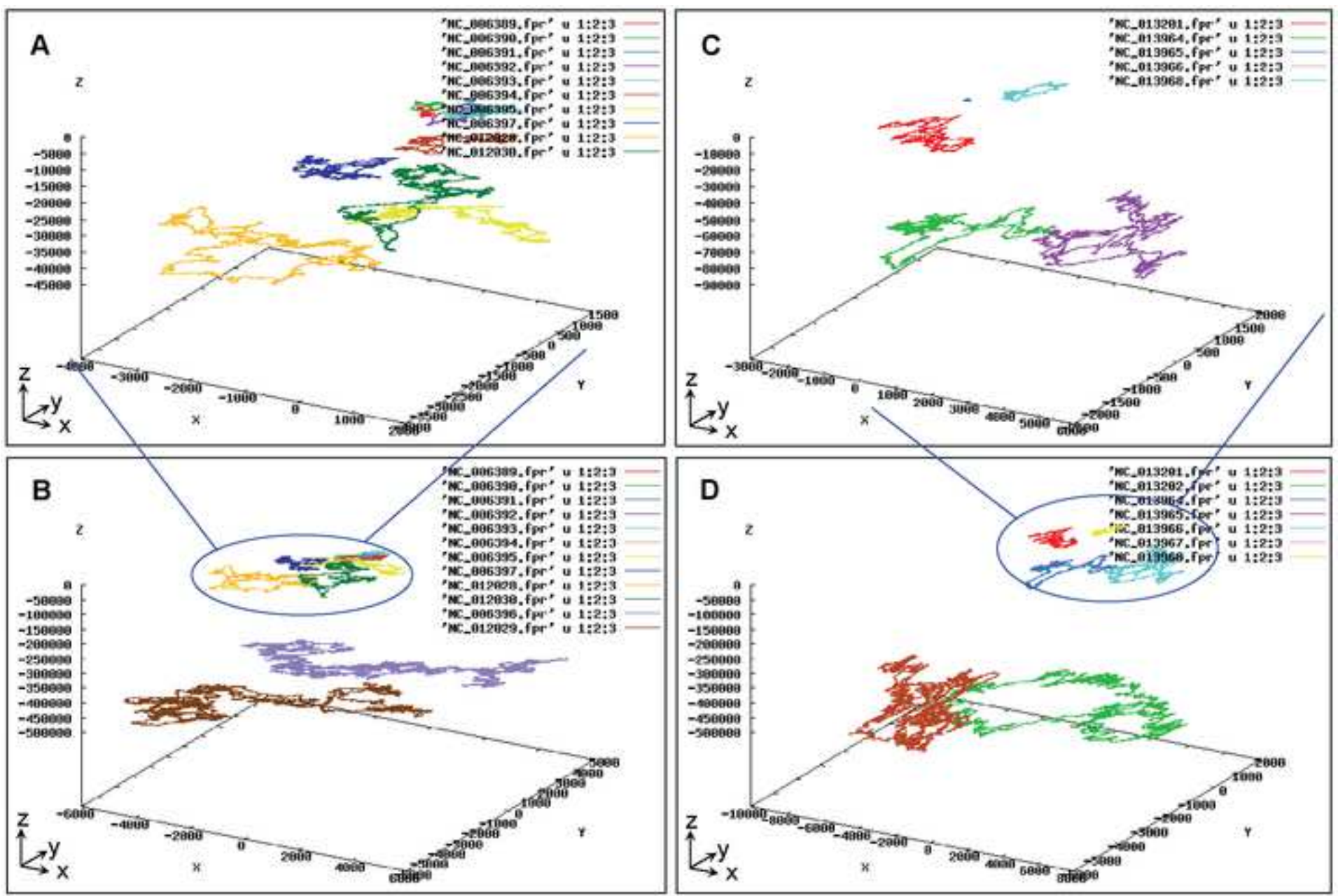

Figure
Click here to download high resolution image

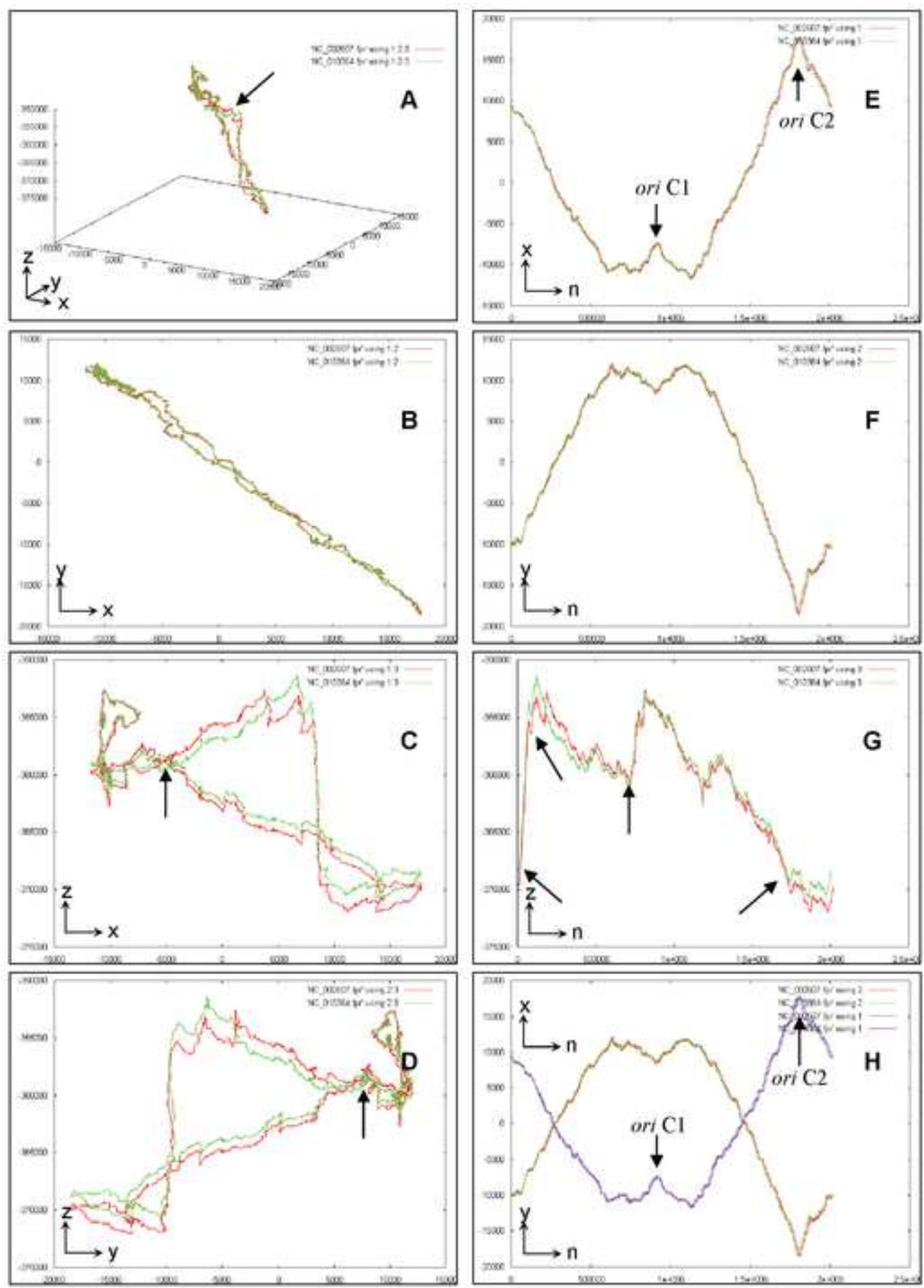

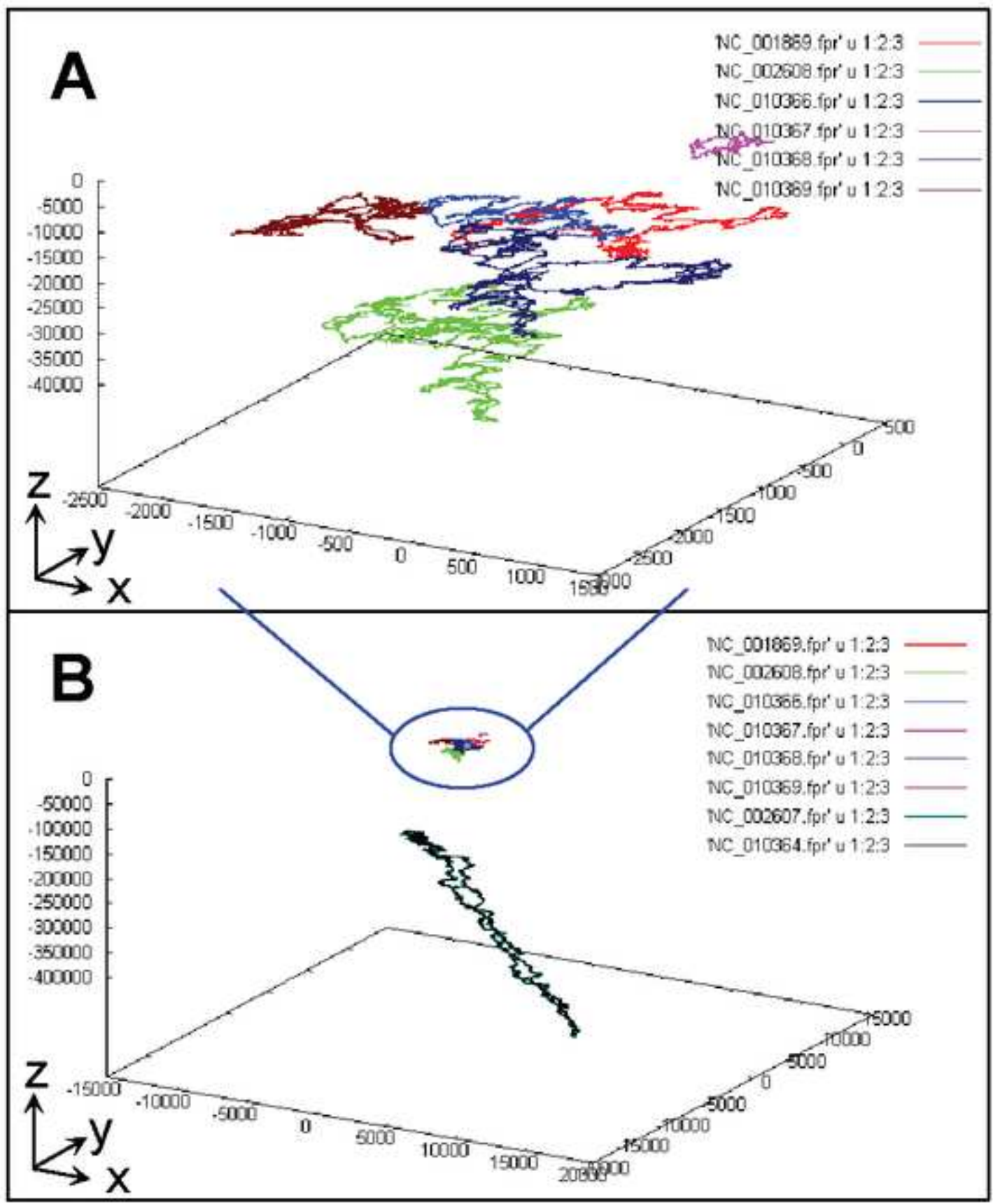

Figure
Click here to download high resolution image

A

B

C

Figure
Click here to download high resolution image

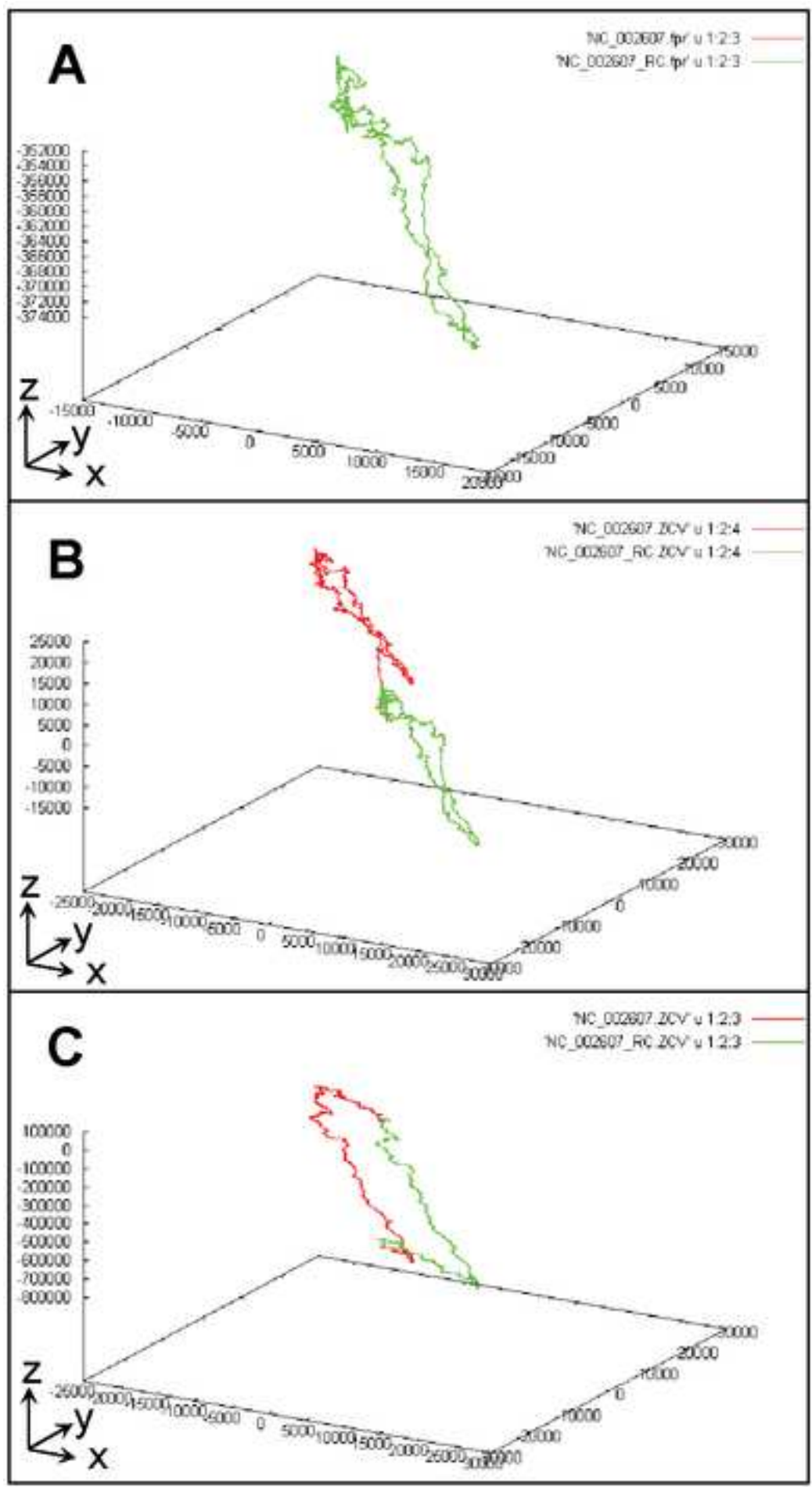